# Legal Judgment Prediction (LJP)
# Amid the Advent of Autonomous AI Legal Reasoning

**Dr. Lance B. Eliot**
Chief AI Scientist, Techbruim; Fellow, CodeX: Stanford Center for Legal Informatics
Stanford, California, USA

**Abstract**

Legal Judgment Prediction (LJP) is a longstanding and open topic in the theory and practice-of-law. Predicting the nature and outcomes of judicial matters is abundantly warranted, keenly sought, and vigorously pursued by those within the legal industry and also by society as a whole. The tenuous act of generating judicially laden predictions has been limited in utility and exactitude, requiring further advancement. Various methods and techniques to predict legal cases and judicial actions have emerged over time, especially arising via the advent of computer-based modeling. There has been a wide range of approaches attempted, including simple calculative methods to highly sophisticated and complex statistical models. Artificial Intelligence (AI) based approaches have also been increasingly utilized. In this paper, a review of the literature encompassing Legal Judgment Prediction is undertaken, along with innovatively proposing that the advent of AI Legal Reasoning (AILR) will have a pronounced impact on how LJP is performed and its predictive accuracy. Legal Judgment Prediction is particularly examined using the Levels of Autonomy (LoA) of AI Legal Reasoning, plus, other considerations are explored including LJP probabilistic tendencies, biases handling, actor predictors, transparency, judicial reliance, legal case outcomes, and other crucial elements entailing the overarching legal judicial milieu.

## 1 Background on Legal Judgment Prediction

In Section 1 of this paper, the literature on Legal Judgment Prediction is introduced and addressed. Doing so establishes the groundwork for the subsequent sections. Section 2 introduces the Levels of Autonomy (LoA) of AI Legal Reasoning (AILR), which is instrumental in the discussions undertaken in Section 3. Section 3 provides an indication of the field of Legal Judgment Prediction as applied to the LoA AILR, along with other vital facets. Section 4 provides various additional research implications and anticipated impacts upon salient practice-of-law considerations.

This paper then consists of these four sections:
- Section 1: Background on Legal Judgment Prediction
- Section 2: Autonomous Levels of AI Legal Reasoning
- Section 3: Legal Judgment Prediction and AI Legal Reasoning
- Section 4: Additional Considerations and Future Research

### 1.1 Overview of Legal Judgment Prediction

Trying to make predictions about judicial matters is an ongoing and longstanding preoccupation that continues to be an open issue in both the theory of the law and the practice of the law [9] [44] [76] [91]. Making predictions about judicial aspects has been difficult and remains unresolved as to how to best or optimally derive such predictions [62] [78]. The advent of computers furthered the effort of making judicial predictions by providing a means of readily undertaking mathematical calculations and computations to aid in rendering predictions. Gradually, as hardware has become faster and less costly, and as software has become more advanced, the use of a myriad of sophisticated statistical techniques and models have also been employed for predicting judicial aspects [51] [54] [55] [77]. Included amongst these computer-based efforts has been the use of Artificial Intelligence (AI) capabilities, such as the use of Natural Language Processing (NLP), Machine Learning (ML), Knowledge-Based Systems (KBS), and the similar AI-enabled technologies and innovations [2] [4] [5] [52] [53] [85].

The focus of making judicial predictions is often referred to as Legal Judgment Prediction (LJP).

For example, Zhong et al [92] describe LJP in this manner: "Legal Judgment Prediction (LJP) aims to predict the judgment result based on the facts of a case and becomes a promising application of artificial intelligence techniques in the legal field. In real-world scenarios, legal judgment usually consists of multiple subtasks, such as the decisions of applicable law articles, charges, fines, and the term of penalty." And, similarly, this indication of the meaning of Legal Judgment Prediction by Xu et al [91]: "Legal judgment prediction (LJP) aims to predict a case's judgment results, such as applicable law articles, charges, and terms of penalty, based on its fact description."

There have been alternative interpretations of the scope associated with the phraseology of *Legal Judgment Prediction*.

Some indicate or infer that the phrasing is akin to a solely outcome-oriented viewpoint, while others use LJP in a wider meaning as encompassing any semblance of judicial decision making. Thus, the moniker could be interpreted as "Legal Outcome Prediction" or it could be viewed as Legal Decision-Making Prediction," a decidedly different connotation. The difference is that in the former case the focus and scope consist of predominantly aiming to predict the judicial result or judicial outcome, only, and doing so without any attention to any intermediary judicial elements, while in the latter case the predictive interest is widely attuned to any form of judicial decision making rather than uniquely an outcomes-based aim.

For purposes of this paper, the latter formulation is taken as the meaning of Legal Judgment Prediction and used in the context of being able to widely predict all manner of judicial judgments, wherein "judgment" does not refer to solely a final decision or outcome but to the presence and use of judicial choices and decision making, for which can occur throughout at any point during a legal case lifecycle and in other judicial contexts too. That being said, it can be acknowledged that this widened option provides a more arduous hurdle for the effort to craft predictive models and render apt predictions. Meanwhile, to be clear, the former interpretation would be considered a subset of the wider definition, and therefore there is nothing somehow inconsistent with the narrower focus (it is consistent in the sense that it provides ample attention to a specialization or narrow subset within the larger set of considerations).

To some, the desire or need to make predictions about judicial matters is patently obvious and straightforward, suggesting that there ought not to be any debate about the intrinsic value in finding ways to make such predictions. The logical argument can be readily made that prediction is part-and-parcel of most everything undertaken in the practice of law, including making predictions about how a legal case will fare, how a judge will decide a legal matter, etc. As succinctly stated by Martin et al [62]: "Legal academics, too, possess expertise that should enable them to forecast legal events with some accuracy. After all, the everyday practice of law requires lawyers to predict court decisions in order to advise clients or determine litigation strategies." This points out that if the practice of law is greatly in need of being able to make judicial predictions, the theory of law and scholars of law would presumably be desirous of aiding in that quest as a means of furthering the underlying matters of law as a body of knowledge and a discipline of academic study.

In short, there does not appear to be an overarching objection per se to the notion of seeking to make judicial predictions and aiming to improve and enhance the capability to make such predictions. That being said, there are concerns expressed about how the act of making judicial predictions can adversely impact the rendering of judicial matters, perhaps telegraphing beforehand choices that then become chosen merely because they were predicted to occur (a self-fulfilling prophecy phenomenon), or that otherwise influence or shape those judicial judgments that have been subject to prediction [19]. Also, there are a wide array of ethical questions that arise, such as whether sophisticated and AI-based LJP models might only be available to those that can afford such mechanizations, therefore arming those that can afford such accouterments while leaving those that cannot afford them a less powerful position of legal armament [20]. These societal and legal ramifications are certainly noteworthy, though they are not the specific purview of this paper, nonetheless, those qualms about the potential adverse consequences of LJP are urged herein as worthy of ongoing research and attention (as mentioned in Section 4).

On a related note, here is a somewhat curt but revealing indication of how the self-fulfilling prophecy could work its way into efforts to undertake Legal Judgement Prediction [20]: "To take a fanciful example that proves the point, suppose an attorney in arguing to a court states that a team of his investigators has been observing the presiding judge every morning. The judge walks out of his downtown apartment and stops for a quick breakfast either at McDonald's or at Dunkin' Donuts. On McDonald's days his decisions favor the plaintiff 84 percent of the time. On Dunkin' days, his decisions favor the defendant 90 percent of the time. The attorney then argues that since he is representing the plaintiff, and since the judge this morning breakfasted at McDonald's, sound principles of statistical sociology require the judge to decide the case in favor of his client."

**1.2 SCOTUS As Judicial Predictions Aim**

One area of Legal Judgment Prediction that has received extensive attention entails making predictions about the Supreme Court of the United States (SCOTUS), logically so, due to fact that the Supreme Court and its several Justices are serving as the highest court in the land and the pinnacle at which laws can arrive for adjudication within the judicial system of the United States.

One of the earliest depicted quantitative analyses involving SCOTUS decision predictions was described by Kort in 1957 [54]: "This study represents an attempt to apply quantitative methods to the prediction of human events that generally have been regarded as highly uncertain, namely, decisions by the Supreme Court of the United States. The study is designed to demonstrate that, in at least one area of judicial review, it is possible to take some decided cases, to identify factual elements that influenced the decisions, to derive numerical values for these elements by using a formula, and then to predict correctly the decisions of the remaining cases in the area specified. The analysis will be made independently of what the Court said by way of reasoning in these cases; it will rely only on the factual elements which have been emphasized by the justices in their opinions and on their votes to affirm or set aside convictions. Changes in Court personnel made no decisive difference in the pattern of judicial action in this area; so the analysis will not need to take into account the fact that twenty-five different justices have occupied the nine seats on the Court during the period covered, *i.e.*, the past quarter century."

Next, Lawlor et al [56] in the early 1960s described how computers were being used to predict SCOTUS decisions, and offered in their paper an indication of specific application to right-to-counsel cases. Keep in mind that computers in that era were relatively slow and less capable than today's computers, and costlier to leverage, yet pointedly were already beginning to be used for LJP efforts. As stated by Grunbaum et al [41] in that same 1960's time period: "Predictions of the outcome of litigation by statistical methods is a relatively new and controversial field of study made possible by computers."

It would seem that almost immediately upon leveraging computers for undertaking Legal Judgment Prediction that debates over the appropriate or best means to construct such predictive models or calculations came to the forefront. For example, Grunbaum et al [41] shared insights into whether the unit of prediction should be the SCOTUS overall outcome or decision, or whether the focal point should be on the Supreme Court Justices per se, and examined numerous studies at that time, reaching this conclusion: "They show that individual Justices, rather than the Court as a whole, should be used as the unit of prediction." Debate and lack of agreement over the "best" unit of prediction for LJP continue to this day, remaining unsettled and undeniably a still unresolved question.

Beyond the question of the unit of prediction, there was also effort involved in trying to ascertain which model or statistical approach might be better suited for Legal Judgment Prediction. Nagel in 1963 proffered this [68]: "This article illustrates and systematically compares three methods for quantitatively predicting case outcomes. The three methods are correlation, regression, and discriminant analysis, all of which involve standard social science research techniques." The focus was about SCOTUS [68]: "The cases used to illustrate the methods consist of 149 civil liberties cases decided by the United States Supreme Court from 1956 through 1960."

The predictor variables of interest by Nagel in the 1963 paper were [68]: "All three methods rely on the relationship between case outcomes and various predictor variables. In this article, the outcome to be predicted is whether a given civil liberties case will be decided in the direction of narrowing civil liberties or in the direction of broadening civil liberties." And, as indicative of that time period, punch cards were used to feed the data into the statistical programs being used [68]: "To use a computer program for regression or discriminant analysis, one punched card per case is needed unless the number of variables necessitates the use of more. Certain columns on each card should be set aside for each variable. Thus, if hole 1 is punched on column 12 of the card corresponding to case 23, this punch might indicate that a certain variable was present."

During the 1980s, SCOTUS predictive models became more pronounced in using several variables as part of the computational approaches, and also emphasized the importance of utilizing the models as both being predictive and potentially being explanatory as to judicial behavior and judgment. This is illustrated via Segal [77]: "The overwhelming consensus of Fourth Amendment scholars is that the Supreme Court's sea and seizure cases are a mess. This article proposes that the confusion arises from the manner in which the cases were studied, not from the decisions themselves. A legal model with variables that include the prior justification of the search, the nature of the intrusion, and a few mitigating circumstance used to explain the Court's decisions on the reasonableness of a given search or seizure. The parameters are estimated through probit. The results show that the search and seizure cases are much more ordered than had commonly been believed. Virtually all of the estimates are as expected. Additionally, the Court is shown to act favorably toward the federal government than toward the states. Preliminary analysis suggests the model has predictive as well as explanatory value."

In more recent times, the SCOTUS predictive models have gradually advanced beyond one-dimensional versions and become multi-dimensional. This is highlighted by the work of Lauderdale and Clark [55]: "One-dimensional spatial models have come to inform much theorizing and research on the U.S. Supreme Court. However, we argue that judicial preferences vary considerably across areas of the law, and that limitations in our ability to measure those preferences have constrained the set of questions scholars pursue. We introduce a new approach, which makes use of information about substantive similarity among cases, to estimate judicial preferences that vary across substantive legal issues and over time. We show that a model allowing preferences to vary over substantive issues as well as over time is a significantly better predictor of judicial behavior than one that only allows preferences to vary over time. We find that judicial preferences are not reducible to simple left-right ideology and, as a consequence, there is substantial variation in the identity of the median justice across areas of the law during all periods of the modern court. These results suggest a need to reconsider empirical and theoretical research that hinges on the existence of a single pivotal median justice."

There has also been keen interest in exploring the use of classification trees as a means to aid in bolstering the Legal Judgment Prediction capacity, once again using SCOTUS as an aim for making judicial predictions, per Kastellec [51]: "A key question in the quantitative study of legal rules and judicial decision making is the structure of the relationship between case facts and case outcomes. Legal doctrine and legal rules are general attempts to define this relationship. This article summarizes and utilizes a statistical method relatively unexplored in political science and legal scholarship—classification trees—that offers a flexible way to study legal doctrine. I argue that this method, while not replacing traditional statistical tools for studying judicial decisions, can better capture many aspects of the relationship between case facts and case outcomes. To illustrate the method's advantages, I conduct classification tree analyses of search and seizure cases decided by the U.S. Supreme Court and confession cases decided by the courts of appeals. These analyses illustrate the ability of classification trees to increase our understanding of legal rules and legal doctrine."

This brings us to the latest variants of studying and attempting to predict SCOTUS, including the work by Katz et al [52]. Making use of an AI-like technique consisting of time-evolving random forest classifiers, here is the nature of their Legal Judgment Prediction efforts: "Building on developments in machine learning and prior work in the science of judicial prediction, we construct a model designed to predict the behavior of the Supreme Court of the United States in a generalized, out-of-sample context. To do so, we develop a time-evolving random forest classifier that leverages unique feature engineering to predict more than 240,000 justice votes and 28,000 cases outcomes over nearly two centuries (1816-2015). Using only data available prior to decision, our model outperforms null (baseline) models at both the justice and case level under both parametric and non-parametric tests. Over nearly two centuries, we achieve 70.2% accuracy at the case outcome level and 71.9% at the justice vote level. More recently, over the past century, we outperform an in-sample optimized null model by nearly 5%. Our performance is consistent with, and improves on the general level of prediction demonstrated by prior work; however, our model is distinctive because it can be applied out-of-sample to the entire past and future of the Court, not a single term. Our results represent an important advance for the science of quantitative legal prediction and portend a range of other potential applications."

Part of the impetus for the approach used by Katz et al was stated as due to faltering or weaknesses in prior works of Legal Judgment Prediction, namely containing these faults or limitations [52]: "Despite the multitude of pundits and vast human effort devoted to the task, the quality of the resulting predictions and the underlying models supporting most forecasts is unclear. Not only are these models not backtested historically, but many are difficult to formalize or reproduce at all. When models are formalized, they are typically assessed ex post to infer causes, rather than used ex ante to predict future cases."

This raises an important point about the assumed underlying principle about much of the Legal Judgment Prediction capacities, pointedly that the primary means of making a prediction is often lacking in being reproducible and that one-instance exemplars do little to showcase true predictive power. In addition, the facet that reliance upon prior data can be detrimental over-reliance [52]: 'Court outcomes are potentially influenced by a variety of dynamics, including public opinion, inter-branch conflict, both changing membership and shifting views of the Justices, and judicial norms and procedures. The classic adage 'past performance does not necessarily predict future results' is very much applicable."

Besides ultimately comparing the predictions of LJP to the actual outcomes, another avenue of gauging the efficacy of LJP models involves the use of legal experts that are asked to make predictions too, as indicated in Martin et al [62]: "Employing two different methods, we attempted to predict the outcome of every case pending before the Supreme Court during its October 2002 term and compared those predictions to the actual decisions. One method used a statistical forecasting model based on information derived from past Supreme Court decisions. The other captured the expert judgments of legal academics and professionals. Using these two distinct methods allows us to test their predictive power not only against actual Court outcomes, but also against each other."

As to the rationale for utilizing both statistical modeling and the polling of legal experts, the basis is depicted this way [62]: "The critical difference between the two methods of prediction lies not in the law/politics dichotomy, but in the nature of the inputs used to generate predictions. The statistical model looked at only a handful of case characteristics, each of them gross features easily observable without specialized training. The legal experts, by contrast, could use particularized knowledge, such as the specific facts of the case or statements by individual justices in similar cases. The statistical model also differed from the experts in explicitly taking into account every case decided by this natural court prior to the 2002 term. No individual could have such comprehensive knowledge of the Court's output for the last eight terms, and so the experts necessarily relied on fewer (albeit more detailed) observations of past Court behavior. Not surprisingly, these different decision-making processes often resulted in divergent predictions in particular cases."

Arising from the advent of online access and the widespread Internet, some assert that Legal Judgment

Prediction can be or should be more than the running of computer-based programs, incorporating the opinions and predictions of human experts about judicial matters, and even going so far as including a larger audience of human predictors via the use of a modern-day crowdsourcing approach [52]: "Future research will seek to find the optimal blend of experts, crowds, and algorithms as some ensemble of these three streams of intelligence likely will produce the best performing model for a wide class of prediction problems."

## 1.3 AI and Legal Judgment Prediction

As earlier indicated, AI has increasingly been utilized for Legal Judgment Prediction, as illustrated via some of the SCOTUS prediction studies mentioned in Section 1.2. Enlarging the viewpoint beyond SCOTUS, AI is being used in numerous ways for a variety of judicial milieu.

For example, in the work by Aletras et al [2], the use of NLP and ML was employed for predicting legal cases of the European Court of Human Rights: "Recent advances in Natural Language Processing and Machine Learning provide us with the tools to build predictive models that can be used to unveil patterns driving judicial decisions. This can be useful, for both lawyers and judges, as an assisting tool to rapidly identify cases and extract patterns which lead to certain decisions. This paper presents the first systematic study on predicting the outcome of cases tried by the European Court of Human Rights based solely on textual content. We formulate a binary classification task where the input of our classifiers is the textual content extracted from a case and the target output is the actual judgment as to whether there has been a violation of an article of the convention of human rights. Textual information is represented using contiguous word sequences, i.e., N-grams, and topics. Our models can predict the court's decisions with a strong accuracy (79% on average). Our empirical analysis indicates that the formal facts of a case are the most important predictive factor. This is consistent with the theory of legal realism suggesting that judicial decision-making is significantly affected by the stimulus of the facts. We also observe that the topical content of a case is another important feature in this classification task and explore this relationship further by conducting a qualitative analysis."

An important point in these studies includes the facet that there are judicial factors at play, along with the need to consider so-called non-legal factors that nonetheless impact the predictive capacity for Legal Judgment Predictions [2]: "These results could be understood as providing some evidence for judicial decision-making approaches according to which judges are primarily responsive to non-legal, rather than to legal, reasons when they decide appellate cases."

When stated bluntly, one viewpoint is that the justices themselves are not merely some form of automata that render legal rulings in a logically and purely mathematically systematic way, instead of as indicated by Grunbaum [40]: "And lastly . . . maybe it should be firstly . .. judges have personalities. They have prejudices and stomach aches and pride and stalled cars and inspirations and hangovers and far visions and sore feet. All judges try, and most succeed in reducing the impact of 'gastronomical jurisprudence,' but few reduce its effect to zero. For after all judges are. . . thanks be to Heaven . . . human. They are not computers controlled by always knowable inputs. Neither are they scientists indifferently imposing inexorable rules. They are only humans, judging only humans, hopefully themselves in turn to be similarly judged."

Per the discussion in Section 1.1, Legal Judgment Prediction can be viewed in a larger context than the sole focus of judicial outcomes, and for which an example of AI used to predict billings or legal fee charges for a given legal case indicates this wide interpretation [59]: "The charge prediction task is to determine appropriate charges for a given case, which is helpful for legal assistant systems where the user input is fact description. We argue that relevant law articles play an important role in this task, and therefore propose an attention-based neural network method to jointly model the charge prediction task and the relevant article extraction task in a unified framework. The experimental results show that, besides providing legal basis, the relevant articles can also clearly improve the charge prediction results, and our full model can effectively predict appropriate charges for cases with different expression styles."

This particular study also raises the point that the source materials used for making Legal Judgment Predictions do not necessarily need to be only legal documents or artifacts, and could very well be potentially "non-legal" materials too, such as news related elements [59]: "By experimenting on news data, we show that, although trained on judgment documents, our model also has reasonable generalization ability on fact descriptions written by non-legal professionals. While promising, our model still cannot explicitly handle multidefendant cases, and there is also a clear gap between our model and the upper bound improvement that relevant articles can achieve."

One concern raised about the use of AI and in particular Machine Learning is that the data used as input is oftentimes required to be pre-labeled, tending to entail manual and labor-intensive efforts to get the data into suitable categorization for use by the ML models. This point is raised by Xu et al [91]: "Existing approaches for legal judgment prediction (LJP) are mainly divided into three categories. In early times, works usually focus on analyzing existing legal cases in specific scenarios with mathematical and statistical algorithms. However, these methods are limited to small datasets with few labels. Later, a number of machine learning-based methods were developed to solve the problem of LJP, which almost combine some manually designed features with a linear classifier to improve the performance of case classification. The shortcoming is that these methods rely heavily on manual features, which suffer from the generalization problem. In recent years, researchers tend to exploit neural networks to solve LJP tasks."

The work by Xu et al [91] proposes an innovative means to cope with these issues: "To solve the confusing charges issue, we propose an end-to-end framework, i.e., Law Article Distillation based Attention Network (LADAN). LADAN uses the difference among similar law articles to attentively extract features from law cases' fact descriptions, which is more effective in distinguishing confusing law articles, and improve the performance of LJP." And as further elaborated [91]: "For an input law case, we learn both macro- and microlevel features. Macro-level features are used for predicting which community includes the applicable law articles. Micro-level features are attentively extracted by the attention vector of the selected community for distinguishing confusing law articles within the same community."

Extending the notion of looking beyond the outcome itself, and thus considering intermediary states of a legal case, the work by Keown [53] provides a methodical means of using a state-of-the-case series of pinpoints or junctures as part of the mathematical modeling for Legal Judgment Prediction, as described this way: "The fact pattern underlying a judicial decision comprises issues that may be classified either as (1) evidence and argument supporting the position of plaintiff denoted by the symbol P, or (2) evidence and argument supporting that of defendant denoted by D. In law, of course, who is plaintiff and who is defendant may depend on which party wins the race to the courthouse, rather than on the nature of the dispute involved."

Furthermore, this [53]: "In civil cases, plaintiff wins his case if the trier of fact, sometimes a judge and sometimes a jury, finds a preponderance of the evidence in his favor. Considering D and P as conflicting factors in a judicial process enjoying a suitably discontinuous behavior, one arrives by means of the Zeemanian process at: Hypothesis I. The standard model of the judicial process is a cusp catastrophe with plaintiffs evidence and argument denoted by P and defendant's evidence and argument denoted by D as conflicting factors determining the outcome." And then emphases the states in time facets [53]: "As intended, these terms indicate that, with certain exceptions to be discussed below, the state of the case is (defined by a point on the behavior manifold that indicates a victory either for plaintiff or for defendant at any given time). If the state of the case is represented by a point on judgment for plaintiff, then it tends to remain there, while if it is represented by a point on judgment for the defendant then by stable equilibrium it tends to remain there. A state of the case in which the outcome is uncertain corresponds to a point on the cross-hatched surface representing a situation of unstable equilibrium. Thus, continued introduction of evidence and further argument soon displaces the state either to judgment for plaintiff or judgment for defendant."

Much of the literature on Legal Judgment Prediction tends to suggest that it is insufficient to make a prediction as though being out-of-the-blue or by some divine means, and argues at times quite vigorously that being able to sensibly or reasonably explain how the prediction was derived is perhaps as important as the prediction itself. Consider the points made by Ashley and Bruninghaus [4] in their efforts to construct their Issue-Based Prediction (IBP) approach: "Computerized algorithms for predicting the outcomes of legal problems can extract and present information from particular databases of cases to guide the legal analysis of new problems. They can have practical value despite the limitations that make reliance on predictions risky for other real-world purposes such as estimating settlement values. An algorithm's ability to generate reasonable legal arguments also is important. In this article, computerized prediction algorithms are compared not only in terms of accuracy, but also in terms of their ability to explain predictions and to integrate predictions and arguments. Our approach, the Issue-Based Prediction algorithm, is a program that tests hypotheses about how issues in a new case will be decided. It attempts to explain away counterexamples inconsistent with a hypothesis, while apprising users of the counterexamples and making explanatory arguments based on them."

As a further extension of this explanatory notion, the IBP was enhanced as SMILE+IBP, as described in [5]: "Work on a computer program called SMILE + IBP (SMart Index Learner Plus Issue-Based Prediction) bridges case-based reasoning and extracting information from texts. The program addresses a technologically challenging task that is also very relevant from a legal viewpoint: to extract information from textual descriptions of the facts of decided cases and apply that information to predict the outcomes of new cases. The program attempts to automatically classify textual descriptions of the facts of legal problems in terms of Factors, a set of classification concepts that capture stereotypical fact patterns that effect the strength of a legal claim, here trade secret misappropriation. Using these classifications, the program can evaluate and explain predictions about a problem's outcome given a database of previously classified cases. This paper provides an extended example illustrating both functions, prediction by IBP and text classification by SMILE, and reports empirical evaluations of each. While IBP's results are quite strong, and SMILE's much weaker, SMILE + IBP still has some success predicting and explaining the outcomes of case scenarios input as texts. It marks the first time to our knowledge that a program can reason automatically about legal case texts."

As also mentioned in Section 1.1, not everyone necessarily concurs that making judicial predictions is appropriate, regardless of how it is undertaken, which recently came to the fore when France banned certain aspects of publishing judge analytics in 2019, as indicated in [84]: "France has banned the publication of judge analytics, and breaking this law carries up to five years in prison. The new Article 33 of the Justice Reform Act reads: 'No personally identifiable data concerning judges or court clerks may be subject to any reuse with the purpose or result of evaluating, analyzing or predicting their actual or supposed professional practices. The violation of this law shall be punished by the measures outlined in articles 226-18, 226-24, and 226-31 of the penal code, without prejudice of the measures and sanctions provided for under the law 78-17 of 6 January 1978 concerning data processing, files and freedoms,' as translated by Rebecca Loescher, a professor of French at St. Edward's University at the Angers, France campus. The law appears to apply to anyone—individuals, researchers, technology companies."

Returning to the point about aiming solely at the outcome of a legal case, it is perhaps confounding to consider exactly what is considered the outcome if there is a possibility that a case might be appealed. In other words, when referring to an outcome, does the outcome imply the true final outcome after all appeals have been exhausted and potentially a final court has ruled or does outcome refer to the initial ruling or judgment about a legal case. Per Atkinson et al [7]: "The point is that the outcome of a case is often not clear: in any serious legal dispute there are opposing arguments, and very often opinions differ as to who has the better of it. Decisions are reversed on appeal, and may be reversed again at the highest level of appeal."

And, once again, theory dovetails into the Legal Judgment Prediction realm and as embodied perhaps in AI efforts as depicted by McCarty [63]: "Using a variety of schemes to represent different kinds of argument (such as Argument from Expert Opinion, Argument from Negative Consequences, Argument from Rules, etc.) was introduced into AI and Law. Argumentation schemes can be seen as a generalisation of the rules of inference." In fact, there is an argument to be made that predictions should be based upon and potentially preceded with a strong theory construction about the law [7]: "Some researchers have argued that reasoning with legal cases should be seen as a process of theory construction, following the ideas of McCarty. The idea is to construct a theory which will explain the past cases and determine an outcome for the current case."

This also potentially recognizes the multitude of subtasks within a legal case and the potential for modeling those subtasks as part of the outcome-focused predictions, utilizing a Directed Acyclic Graph approach [92]: "While most existing works only focus on a specific subtask of judgment prediction and ignore the dependencies among subtasks, we formalize the dependencies among subtasks as a Directed Acyclic Graph (DAG) and propose a topological multi-task learning framework, TOPJUDGE, which incorporates multiple subtasks and DAG dependencies into judgment prediction. We conduct experiments on several real-world large-scale datasets of criminal cases in the civil law system. Experimental results show that our model achieves consistent and significant improvements over baselines on all judgment prediction tasks."

As indicative of the legal case lifecycle and the role of Legal Judgment Prediction, consider these types of questions that are customarily asked about a legal case [78]: "Usually, the process of prediction starts with one or more questions. Whether to take a case in hand or not? Whether to settle the case outside or take it to the court? Will the settlement amount be worth it? What are the chances of winning the case? These are some of the questions that involve predicting the outcome of a case and the legal practitioners have to deal with, on a regular basis. These questions represent the importance of outcome prediction in case selection, making settlement decisions, and various aspects of legal processes."

Consider too that there are judicial per se factors and then there are extra-judicial factors to be encompassed when performing LJP [78]: "Considering the fact that judges are human beings, this approach assumes that judges may be ideologically inclined towards some side in various issues, have their own perception and other biases, and various other social and political factors that affect their judgment such as mental resources of a judge and decision of lower court. Hence, descriptors considered for predicting outcomes are extra-judicial factors that may generate or represent human bias. Examples of these factors are votes of other justices, justice gender, case origin, petitioner type, respondent type, the ideological direction of the court, etc."

Besides conventional numeric coding of judicial factors, there is also a linguistic approach that can be utilized for Legal Judgment Prediction, as indicated in [78]: "Another approach considers the linguistic features of legal judgments for predicting outcomes. Ngo tried to predict outcomes on a database of 2019 Dutch legal cases according to their linguistic features. Some of the considered features were word count and frequency of different types of pronouns. In, the authors replaced case-specific names and instances by their role and used propositional patterns for predicting trade secret cases."

For those seeking to establish law as a form of science, one that would adhere to rigorous principles of nature and be readily described via formulas and the means of science, and thus presumably be more readily predicted and predictable, Noonan offers this thought in his 1961 work about the law [69]: "Other sciences may be more easily circumscribed. Their subject matter is defined by the Baconian purpose for which they are normally pursued: the prediction and control of the properties of a particular kind of physical matter. It would appear that a similar ambition to achieve prediction and control 'like a science' may have animated such Austinian offshoots as Legal Realism."

**1.4 Theories of Law and LJP**

Embroiled within Legal Judgment Prediction is the role of underlying theories about the law.

In particular, there has been much focus on the emergence of legal realism versus legal formalism, and thus this carries over inevitably into the predictive realm too [2]: "Without going into details with respect to a particularly complicated debate that is out of the scope of this paper, we may here simplify by observing that since the beginning of the 20th century, there has been a major contention between two opposing ways of making sense of judicial decision-making: legal formalism and legal realism. Very roughly, legal formalists have provided a *legal model* of judicial decision-making, claiming that the law is rationally determinate: judges either decide cases deductively, by subsuming facts under formal legal rules or use more complex legal reasoning than deduction whenever legal rules are insufficient to warrant a particular outcome. On the other hand, legal realists have criticized formalist models, insisting that judges primarily decide appellate cases by responding to the stimulus of the facts of the case, rather than on the basis of legal rules or doctrine, which are in many occasions rationally indeterminate."

Seeking to encompass these underlying tensions of legal theories, a study by Hall and Wright utilized content analysis, whose roots can be said to epistemologically be in the legal realism realm, as described in [44]: "To provide methodological guidance, we survey the questions that legal scholars have tried to answer through content analysis, and use that experience to generalize about the strengths and weaknesses of the technique compared with conventional interpretive legal methods. The epistemological roots of content analysis lie in legal realism. Any question that a lawyer might ask about what courts say or do can be studied more objectively using one of the four distinct components of content analysis: 1) replicable selection of cases; 2) objective coding of cases; 3) counting case contents for descriptive purposes; or 4) statistical analysis of case coding."

The researchers point out that there are noted objections to such an approach [44]: "Each of these components contributes something of unique epistemological value to legal research, yet at each of these four stages, some legal scholars have objected to the technique. The most effective response is to recognize that content analysis does not occupy the same epistemological ground as conventional legal scholarship. Instead, each method renders different kinds of insights that complement each other, so that, together, the two approaches to understanding caselaw are more powerful that either alone. Content analysis is best used when each decision should receive equal weight, that is, when it is appropriate to regard the content of opinions as generic data."

The argument for content analysis is further made [44]: "Scholars have found that it is especially useful in studies that question or debunk conventional legal wisdom. Content analysis also holds promise in the study of the connections between judicial opinions and other parts of the social, political, or economic landscape. The strongest application is when the subject of study is simply the behavior of judges in writing opinions or deciding cases. Then, content analysis combines the analytical skills of the lawyer with the power of science that comes from articulated and replicable methods."

As earlier emphasized in Section 1.1, there can be problematic issues underlying the approaches used for Legal Judgment Prediction, including this aspect highlighted about the use of content analysis [44]: "However, analyzing the cause-and-effect relationship between the outcome of cases and the legally relevant factors presented by judges to justify their decisions raises a serious circularity problem. Therefore, content analysis is not an especially good tool for helping lawyers to predict the outcome of cases based on real-world facts. This article also provides guidance on the best practices for using this research method. We identify techniques that meet standards of social science rigor and account for the practical needs of legal researchers. These techniques include methods for case sampling, coder training, reliability testing, and statistical analysis."

## 1.5 Holmes And The Path Of The Law

It would seem that any in-depth discussion about prediction and the law, which entails any notable theoretical semblance, would be inclined or perhaps obligated to address the now-classic "prediction theory of the law" as commonly prompted by the work of Oliver Wendell Holmes in his 1897 "The Path of the Law" work [40]. For those not familiar with the premise, Holmes could be construed as arguing that law is to its essence a form of a prediction.

This is worthwhile for elucidation and detailed consideration in these matters on Legal Judgment Prediction.

A singular quote of Holmes has been promulgated time and again over the years, of which there is a multitude of interpretations about the purported meaning, that quote being this one [40]: 'The prophecies of what the courts will do in fact, and nothing more pretentious, are what I mean by the law."

Before examining the myriad of interpretations, it is instructive to consider the overall context of what else Holmes indicated and how his famous article arrived at that particular statement or sentiment.

First, Holmes indicates that law is weighed in the minds of the public as to the amount of governmental force that can be applied to ensure that the law itself is abided by, such that [40]: "People want to know under what circumstances and how far they will run the risk of coming against what is so much stronger than themselves, and hence it becomes a business to find out when this danger is to be feared. The object of our study, then, is prediction, the prediction of the incidence of the public force through the instrumentality of the courts."

In that frame of thinking, laws can be considered prophecies or predictions about how the ax, as it were, might fall upon someone [40]: "The means of the study are a body of reports, of treatises, and of statutes, in this country and in England, extending back for six hundred years, and now increasing annually by hundreds. In these sibylline leaves are gathered the scattered prophecies of the past upon the cases in which the axe will fall. These are what properly have been called the oracles of the law. Far the most important and pretty nearly the whole meaning of every new effort of legal thought is to make these prophecies more precise, and to generalize them into a thoroughly connected system."

To reiterate then, in Holmes own words, the law is a prediction of what can befall those that possibly avert the law [40]: "But, as I shall try to show, a legal duty so called is nothing but a prediction that if a man does or omits certain things he will be made to suffer in this or that way by judgment of the court; and so of a legal right." And to add emphasis, he states: "I wish, if I can, to lay down some first principles for the study of this body of dogma or systematized prediction which we call the law, for men who want to use it as the instrument of their business to enable them to prophesy in their turn, and, as bearing upon the study, I wish to point out an ideal which as yet our law has not attained."

This then leads to the famous line, which is shown herein in the context of both the preceding background and the entire passage of which the line occurs (it is the last sentence here) [40]: "Take the fundamental question, What constitutes the law? You will find some text writers telling you that it is something different from what is decided by the courts of Massachusetts or England, that it is a system of reason, that it is a deduction from principles of ethics or admitted axioms or what not, which may or may not coincide with the decisions. But if we take the view of our friend the bad man we shall find that he does not care two straws for the axioms or deductions, but that he does want to know what the Massachusetts or English courts are likely to do in fact. I am much of this mind. The prophecies of what the courts will do in fact, and nothing more pretentious, are what I mean by the law."

As an aside, some object to the characterization of "bad man" or "bad men" to which the law is apparently aimed (in today's terminology it would be bad people, which is likely what Holmes had intended), and per Hart, the pointed question is asked [46]: "Why should not law be equally if not more concerned with the 'puzzled man' or 'ignorant man' who is willing to do what is required, if only he can be told what it is? Or with the 'man who wishes to arrange his affairs' if only he can be told how to do it?" This aside is not addressed in this paper herein and merely noted as one of various criticisms or analyses that have been made of Holmes's statements and philosophy about the law.

Continuing, what is perhaps equally crucial in the viewpoint expressed by Holmes is that it is not merely the words of the law that are crucial, but also and perhaps as much so (or more) the operationalizing of the law too [40]: "You see how the vague circumference of the notion of {legal} duty shrinks and at the same time grows more precise when we wash it with cynical acid and expel everything except the object of our study, the operations of the law."

As a point that has become commonly cited about the law presumably (possibly) not being amenable to logic, and perhaps not amenable to mathematical modeling and computations that might attempt to embody or utilize it for predictive purposes, here's what Holmes stated [40]: "So in the broadest sense it is true that the law is a logical development, like everything else. The danger of which I speak is not the admission that the principles governing other phenomena also govern the law, but the notion that a given system, ours, for instance, can be worked out like mathematics from some general axioms of conduct.

Furthermore, the law can turn on a dime, one might assert, and presumably undercut the notion of the past being used to predict the future [40]: "Such matters really are battle grounds where the means do not exist for the determinations that shall be good for all time, and where the decision can do no more than embody the preference of a given body in a given time and place. We do not realize how large a part of our law is open to reconsideration upon a slight change in the habit of the public mind. No concrete proposition is self-evident, no matter how ready we may be to accept it, not even Mr. Herbert Spencer's 'Every man has a right to do what he wills, provided he interferes not with a like right on the part of his neighbors.'"

There is though a role, an important one, in applying overarching statistics and economic postulates to the prediction of the predictions: "The rational study of law is still to a large extent the study of history. History must be a part of the study, because without it we cannot know the precise scope of rules which it is our business to know. It is a part of the rational study, because it is the first step toward an enlightened skepticism, that is, towards a deliberate reconsideration of the worth of those rules. When you get the dragon out of his cave on to the plain and in the daylight, you can count his teeth and claws, and see just what is his strength. But to get him out is only the first step. The next is either to kill him, or to tame him and make him a useful animal. For the rational study of the law the blackletter man may be the man of the present, but the man of the future is the man of statistics and the master of economics. It is revolting to have no better reason for a rule of law than that so it was laid down in the time of Henry IV. It is still more revolting if the grounds upon which it was laid down have vanished long since, and the rule simply persists from blind imitation of the past." Along with this point too [40]: "The statistics of the relative increase of crime in crowded places like large cities, where example has the greatest chance to work, and in less populated parts, where the contagion spreads more slowly, have been used with great force in favor of the latter view. But there is weighty authority for the belief that, however this may be, 'not the nature of the crime, but the dangerousness of the criminal, constitutes the only reasonable legal criterion to guide the inevitable social reaction against the criminal.'"

Out of this array of vital considerations about the law has become an assertion that these points can arrive at a contention that the courts and judges are ultimately the true determiners of what the law is.

Rather than being in the role of law-applier, which in theory the courts and judges are supposed to be focused and limited to thereof, it has been interpreted that Holmes was arguing that the courts and the judges are instead the lawmakers, in lieu of the legislature that was presumed to be the lawmakers. This is argued under the assertion that the courts and the judges are the final arbiters of the meaning of the law, along with the applying of the forces that Holmes argued are the basis for why the laws are seen as having teeth and thus the foundational face for the public to consider when abiding by the laws.

If a law is a predictor of the potential force to be applied, but the courts and the judges can sway that "law" in whatever manner so desired, the law itself no longer is as meaningful as is the judgment that the courts and judges will make, regardless potentially of what the law otherwise seemingly appears to signify. Thus, this leads to the logical conclusion that the focus for Legal Judgment Prediction should not be to the law itself, being a mere secondary or surrogate at best, but instead to the courts and judges as to their judgments (and potential whims, some would argue). As described by D'Amato [20]: "The judge who finds it more interesting to invent new law rather than restate the old is stealthily undermining public confidence in the rule of law and narrowing the ambit of personal freedom. He is acting as a legislator, not a judge—a legislator of the worst sort, who enacts new law and holds it against innocent people who were dutifully complying with the old law. If the judge in addition believes that he embodies the law, he is saying that, if the public wants to understand the law, they should study him. For in the end he has no theory of law. He cannot explain what the law is; he can only say that law is what he does—but he does not say that it is a shortcoming on his part that he cannot explain what the law is. Instead it is a failing on the law's part!"

Indeed, this active role of lawmaking can be construed as a form of reverse legislation [20]: "A judge who invents a rule and applies it retroactively to conduct that has already occurred seems to be engaging in a kind of reverse legislation. If in doing so the judge changes the law retroactively, then indeed it would be reverse-perverse legislation-penalizing a party for failure to obey a rule that the judge has just invented. But if the judge just finds the law as it existed when the facts of the case arose, then the judge is not making new law but only taking a picture of the old— the same picture the litigants could have taken when their case arose."

Returning to the Holmes remark about the law as a prophecy, subsequent to Holmes points becoming popularized, many interpreted that this meant that lawyers are essentially like weather predictors [20]: "After hearing Holmes speak, the scholars and practitioners in the audience undoubtedly understood him to be comparing a lawyer with other public predictors of events such as weather forecasters. This particular comparison has become the standard interpretation of Holmes's prediction theory. Thus a lawyer predicts judicial decisions (which constitute the law), and the meteorologist predicts tomorrow's weather."

Per D'Amato [20], this opened a can of worms amid the legal profession: "It may be warranted to say that legal realism was a disastrous setback for American law. It seemed to justify as an uncontestable fact of empiricism that judges may make all kinds of decisions based upon a wide range of factors: emotions, prejudices (unless they amount to a conflict of interest), party affiliation, rewarding campaign contributors, facile study of the law, liking or disliking the attorneys arguing a case, mere whim, and other bells and whistles. Law-school curriculums are then skewed to prepare students to argue successfully before judges who may only care minimally about what the law says." Leading to this rather stark conclusion about the path of lawyers in the law [20]: "Better yet, once he becomes a judge, he will not have to pay much attention to what the lawyers say about the law (any more than he did in the classroom). For the 'law' will be whatever he proclaims it to be."

In considering the role of uncertainty, which naturally flows from the topic of making predictions (this will be future explored in Section 3), it is insightful to consider that if the law is predicated on the courts and the ruling of judges, the public then would presumably be keenly focused on the present probability of the predicted outcome of the court and the judges, more so than the law per se [20]: "In short, it is the present probability of what the court will do that is of great interest to the bad man rather than the future fact of what the court will decide because by then it will be too late to influence the bad man's decision in the present."

As a vivid illustration of this point about assessing the present probability of a future predicted event, D'Amato offers this illustration [20]: "To this Holmes may have added that all decisions, not just those with legal consequences, are based on probabilities. For example, a jaywalker decides to cross the street. He figures that his chances of reaching the other side are 99.99% He has allowed 0.01% for the possibility that he will stumble halfway through and be run over by a passing car. On top of that calculation he must also consider the probability of being arrested for jaywalking—an arrest that would defeat his purpose of crossing the street in between the traffic signals. He looks around and does not see a police officer. Nevertheless, he has to allocate some degree of probability that a police officer may be standing behind a truck parked on the other side of the street. Thus his probability of successfully jaywalking reduces from 99.99% to, say, 97%. This probability may or may not lead him to decide to jaywalk. There may be other factors that enter into his decision. Every one of those factors can be measured as a numerical probability. His decision whether or not to jaywalk, just like every other decision he makes or will ever make, is based on the summation of all the relevant probabilities of which he is aware at the time he makes his decision."

In short, this aptly sums up the viewpoint about the role of prediction and the law [20]:

"We cannot know exactly what the law is right now when we want to factor it into our decision whether to act or not to act, but we can assign a numerical prediction in the present to what a court will later decide and treat the prediction as the law."

This discussion about Holmes perhaps makes abundantly clear that the act of prediction and the role of Legal Judgment Prediction have merit, substantially so, and presumably can be said to be at the core of the law, including both the theory of the law and the practice of the law.

For those that at times suggest that Legal Judgment Prediction only needs to consider the courts and the judges, omitting entirely the law itself, under the guise that the law will be whatever the courts say it to be, this seems somewhat misguided as a rule of thumb. The law still is nonetheless the essence of what is relied upon, and for which the courts and the judges will presumably emanate from, and thus acting as though for LJP purposes that the laws are unworthy of inclusion seems an extreme and relatively imprudent posture. In that same view, perhaps, taking the extreme of only relying upon the laws as written would seem to be missing the boat, as it were, neglecting to take into account the role and impact of the decisions made by the courts and the judges. All told, attempting to portray the law and the courts/judges as somehow mutually exclusive when undertaking LJP seems exceedingly ill-advised.

In any case, the Holmes exploration has also highlighted the importance of realizing that predictions via Legal Judgment Prediction are not to be cast as imbuing absolute certainty and that there needs to be the realization and infusion of probabilities in order to consider the predictions of proper value and utility.

## 2 Autonomous Levels of AI Legal Reasoning

In this section, a framework for the autonomous levels of AI Legal Reasoning is summarized and is based on the research described in detail in Eliot [27] [28] [29] [30] [31] [32].

These autonomous levels will be portrayed in a grid that aligns with key elements of autonomy and as matched to AI Legal Reasoning. Providing this context will be useful to the later sections of this paper and will be utilized accordingly.

The autonomous levels of AI Legal Reasoning are as follows:

Level 0: No Automation for AI Legal Reasoning

Level 1: Simple Assistance Automation for AI Legal Reasoning

Level 2: Advanced Assistance Automation for AI Legal Reasoning

Level 3: Semi-Autonomous Automation for AI Legal Reasoning

Level 4: Domain Autonomous for AI Legal Reasoning

Level 5: Fully Autonomous for AI Legal Reasoning

Level 6: Superhuman Autonomous for AI Legal Reasoning

### 2.1 Details of the LoA AILR

See **Figure A-1** for an overview chart showcasing the autonomous levels of AI Legal Reasoning as via columns denoting each of the respective levels.

See **Figure A-2** for an overview chart similar to Figure A-1 which alternatively is indicative of the autonomous levels of AI Legal Reasoning via the rows as depicting the respective levels (this is simply a reformatting of Figure A-1, doing so to aid in illuminating this variant perspective, but does not introduce any new facets or alterations from the contents as already shown in Figure A-1).

#### 2.1.1 Level 0: No Automation for AI Legal Reasoning

Level 0 is considered the no automation level. Legal reasoning is carried out via manual methods and principally occurs via paper-based methods.

This level is allowed some leeway in that the use of say a simple handheld calculator or perhaps the use of a fax machine could be allowed or included within this Level 0, though strictly speaking it could be said that any form whatsoever of automation is to be excluded from this level.

#### 2.1.2 Level 1: Simple Assistance Automation for AI Legal Reasoning

Level 1 consists of simple assistance automation for AI legal reasoning.

Examples of this category encompassing simple automation would include the use of everyday computer-based word processing, the use of everyday computer-based spreadsheets, the access to online legal documents that are stored and retrieved electronically, and so on.

By-and-large, today's use of computers for legal activities is predominantly within Level 1. It is assumed and expected that over time, the pervasiveness of automation will continue to deepen and widen, and eventually lead to legal activities being supported and within Level 2, rather than Level 1.

### 2.1.3 Level 2: Advanced Assistance Automation for AI Legal Reasoning

Level 2 consists of advanced assistance automation for AI legal reasoning.

Examples of this notion encompassing advanced automation would include the use of query-style Natural Language Processing (NLP), Machine Learning (ML) for case predictions, and so on.

Gradually, over time, it is expected that computer-based systems for legal activities will increasingly make use of advanced automation. Law industry technology that was once at a Level 1 will likely be refined, upgraded, or expanded to include advanced capabilities, and thus be reclassified into Level 2.

### 2.1.4 Level 3: Semi-Autonomous Automation for AI Legal Reasoning

Level 3 consists of semi-autonomous automation for AI legal reasoning.

Examples of this notion encompassing semi-autonomous automation would include the use of Knowledge-Based Systems (KBS) for legal reasoning, the use of Machine Learning and Deep Learning (ML/DL) for legal reasoning, and so on.

Today, such automation tends to exist in research efforts or prototypes and pilot systems, along with some commercial legal technology that has been infusing these capabilities too.

### 2.1.5 Level 4: Domain Autonomous for AI Legal Reasoning

Level 4 consists of domain autonomous computer-based systems for AI legal reasoning.

This level reuses the conceptual notion of Operational Design Domains (ODDs) as utilized in the autonomous vehicles and self-driving cars levels of autonomy, though in this use case it is being applied to the legal domain [22] [23] [24]. Essentially, this entails any AI legal reasoning capacities that can operate autonomously, entirely so, but that is only able to do so in some limited or constrained legal domain.

### 2.1.6 Level 5: Fully Autonomous for AI Legal Reasoning

Level 5 consists of fully autonomous computer-based systems for AI legal reasoning.

In a sense, Level 5 is the superset of Level 4 in terms of encompassing all possible domains as per however so defined ultimately for Level 4. The only constraint, as it were, consists of the facet that the Level 4 and Level 5 are concerning human intelligence and the capacities thereof. This is an important emphasis due to attempting to distinguish Level 5 from Level 6 (as will be discussed in the next subsection)

It is conceivable that someday there might be a fully autonomous AI legal reasoning capability, one that encompasses all of the law in all foreseeable ways, though this is quite a tall order and remains quite aspirational without a clear cut path of how this might one day be achieved. Nonetheless, it seems to be within the extended realm of possibilities, which is worthwhile to mention in relative terms to Level 6.

### 2.1.7 Level 6: Superhuman Autonomous for AI Legal Reasoning

Level 6 consists of superhuman autonomous computer-based systems for AI legal reasoning.

In a sense, Level 6 is the entirety of Level 5 and adds something beyond that in a manner that is currently ill-defined and perhaps (some would argue) as yet unknowable. The notion is that AI might ultimately exceed human intelligence, rising to become superhuman, and if so, we do not yet have any viable indication of what that superhuman intelligence consists of and nor what kind of thinking it would somehow be able to undertake.

Whether a Level 6 is ever attainable is reliant upon whether superhuman AI is ever attainable, and thus, at this time, this stands as a placeholder for that which might never occur. In any case, having such a placeholder provides a semblance of completeness, doing so without necessarily legitimatizing that superhuman AI is going to be achieved or not. No such claim or dispute is undertaken within this framework.

# 3 Legal Judgement Prediction and AI Legal Reasoning

In this Section 3, various aspects of Legal Judgment Prediction (LJP) will be identified and discussed with respect to AI Legal Reasoning (AILR). A series of diagrams and illustrations are included to aid in depicting the points being made. In addition, the material draws upon the background and LJP research literature indicated in Section 1 and combines with the material outlined in Section 2 on the Levels of Autonomy (LoA) of AI Legal Reasoning.

## 3.1 LJP and LoA AILR

The nature and capabilities of Legal Judgment Prediction will vary across the Levels of Autonomy for AI Legal Reasoning.

Refer to **Figure B-1**.

As indicated, the Legal Judgment Prediction becomes increasingly more sophisticated and advanced as the AI Legal Reasoning increases in capability. To aid in typifying the differences between each of the Levels in terms of the incremental advancement of LJP, the following phrasing is used:
- Level 0: **n/a**
- Level 1: **Rudimentary Calculative**
- Level 2: **Complex Statistical**
- Level 3: **Symbolic Intertwined**
- Level 4: **Domain Predictive**
- Level 5: **Holistic Predictive**
- Level 6: **Pansophic Predictive**

Briefly, each of the levels is described next.

At Level 0, there is an indication of "n/a" at Level 0 since there is no AI capability at Level 0 (the *No Automation* level of the LoA).

At Level 1, the LoA is *Simple Assistance Automation* and this can be used to undertake Legal Judgment Prediction though it is rated or categorized as being rudimentary and making use of relatively simplistic calculative models and formulas. Thus, this is coined as "Rudimentary Calculative."

At Level 2, the LoA is *Advanced Assistance Automation* and the LJP is coined as "Complex Statistical," which is indicative of Legal Judgment Prediction being performed in a more advanced manner than at Level 1. This consists of complex statistical methods such as those techniques mentioned in Section 1 of this paper. To date, most of the research and practical use of Legal Judgment Prediction has been within Level 2. Future efforts are aiming at Level 3 and above.

At Level 3, the LoA is *Semi-Autonomous Automation* and the LJP is coined as "Symbolic Intermixed," which can undertake Legal Judgment Prediction at an even more advanced capacity than at Level 2. Recall, in Level 2, the focus tended to be on traditional numerical formulations for LJP, albeit sophisticated in the use of statistical models. In Level 3, the symbolic capability is added and fostered, including at times acting in a hybrid mode with the conventional numerical and statistical models. Generally, the work at Level 3 to-date has primarily been experimental, making use of exploratory prototypes or pilot efforts.

At Level 4, the LoA is *AILR Domain Autonomous* and coined as "Domain Predictive," meaning that this can be used to perform Legal Judgment Predictions within particular specialties of domains or subdomains of the legal field, but does not necessarily cut across the various domains and is not intended to be able to do so. The predictive capacity is done in a highly advanced manner, incorporating the Level 3 capabilities, along with exceeding those levels and providing a more fluent and capable predictive means.

At Level 5, the LoA is *AILR Fully Autonomous* and coined as "Holistic Predictive," meaning that the use of Legal Judgment Predictions can go across all domains and subdomains of the legal field. The predictive capacity is done in a highly advanced manner, incorporating the Level 4 capabilities, along with exceeding those levels and providing a more fluent and capable predictive means.

At Level 6, the LoA is *AILR Superhuman Autonomous*, which as a reminder from Section 2 is not a capability that exists and might not exist, though it is included as a provision in case such a capability is ever achieved. In any case, the Legal Judgment Prediction at this level is considered "Pansophic Predictive" and would encapsulate the Level 5 capabilities, and then go beyond that in a manner that would leverage the AI superhuman capacity.

## 3.2 Legal Judgment Prediction: Approaches Utilized

Based on the discussion in Section 1, it is useful to consider the overarching nature of the approaches utilized in ascertaining Legal Judgment Prediction efforts.

Refer to **Figure B-2**.

As indicated, within the overall legal milieu there are the laws and the courts, of which there are then legal outcomes that are derived and produced. Presumably, the laws are made by lawmakers, while the courts consist of judges in the role of law-appliers (see Section 1 for further explanation about these matters).

For any given specific legal case, the goal of Legal Judgment Prediction is ostensibly to predict the legal outcome of that particular legal case.

The predictive approach typically uses as key factors:
- Case-Specific Legal Factors
- Judge-Specific Legal Factors
- Exogenous Factors
- Other

In the case-specific legal factors, the data or information collected, assessed, and used for making the outcome prediction is often quantified numerically. As per the discussion in Section 1, this might include codifying the factors into binary numeric values, stratified ranges, and so on. Also, there are linguistic bases for conducting the predictive efforts, involving examining the linguistical nuances and nature of the case. There is also the symbolic understanding approach that attempts to go beyond the traditional numeric or quantified approaches and seemingly include some form of comprehension toward the symbolic nature of the elements involved to make predictions.

Per the points made in Section 3, the judge-specific factors can also be included in the effort to undertake Legal Judgment Predictions. These can be categorized into two major sets, consisting of judicial oriented factors (such as the prior rulings of the judge) and extra-judicial oriented factors (e.g., what the judge ate for breakfast, which is an oft used exemplar).

Exogenous factors consist of elements that are beyond those of the case-specific and judge-specific, such as considering societal matters, socio-economic aspects, and the like.

Consider next some variants and additional considerations about the various factors and their usage.

As indicated in **Figure B-3**, for those that argue that judges are apt to be "lawmakers" rather than solely acting as law-appliers, it would imply that the judge-specific factors become even more predominant and crucial than otherwise might be the case.

To further illustrate this perspective about Legal Judgment Prediction, consider the illustrations shown in **Figure B-4**.

All else being equal, it might be the case that the focus of the Legal Judgment Prediction would be based primarily on the laws themselves and less so than the courts, assuming that the courts are considered as consisting of law-appliers rather than as lawmakers. For the viewpoint that the courts and judges are actually working in a lawmaking capacity, presumably, the focus of Legal Judgment Prediction would center on the judges, more so than the laws themselves.

## 3.3 Four-Square Grid of Judicial Role

The discussion in Section 3.2 can be further extended.

Refer to **Figure B-5**.

This depicts a grid consisting of four quadrants.

Along the horizontal axis, there are two ways of interpreting the role of the judiciary, either as strictly law-appliers or as lawmakers. On the vertical axis, there are two ways shown that the law itself might be construed, either consisting of resolute law (as devised by legislative lawmakers) or as a kind of legal scaffolding (meaning that the law is malleable and porous).

The two horizontal and two vertical indications allow for a four-square quadrant representation and thus the grid as depicted.

In theory, the notion of judges as strictly being law-appliers can be mated with the notion of a resolute law facet, and thus the upper left quadrant that is labeled as V1 and indicated as "The Conventional Assumption." This square consists of the perspective that the courts and judges act to operationalize the law, rather than to make law per se.

Considered a more contemporary view is the square labeled as V2, in the lower right quadrant of the grid. This is the veritable "law is what the courts say it is" positioning, consisting of the judges as lawmakers and the law being perceived or acted upon as though it is malleable and porous.

This leaves two additional squares of the quadrant.

The lower left of the grid consists of the judges as lawmakers which is mated with the resolute law notion. As stated, this generally creates an inherent conflict, due to the possibility of essentially having two cooks in the kitchen, as it were.

Finally, the upper right grid indicates the judges as strictly law-appliers when mated with the law as scaffolding. It can be argued that this creates the possibility of a legal void in the law. Inevitably, there are likely pressures that would come to bear as to seeking to close or narrow the void. The intrinsic tension is whether the courts and judges are the appropriate void fillers, or whether the lawmakers of the legislative are the appropriate void fillers, or some combination thereof.

**3.4 Legal Case Lifecycle Aspects**

An important aspect of Legal Judgment Prediction consists of the legal case timeline and the stage at which a prediction is being rendered and the target stage of the prediction.

Much of the research described in Section 1 was generally based on the assumption that a prediction is to be rendered at the start of a case and the target is the outcome of the case. This though is somewhat amorphous and ill-specified and can be reconsidered more comprehensively by considering the full lifecycle of a legal case.

Refer to **Figure B-6**.

Consider these key junctures in a legal case:
- Pre-Case
- Case Start
- Case Underway
- Case Outcome (initial)
- Case Appeal #1
- Case Appeal #N
- Case Outcome (final)

It could be that before a legal case even gets started, there is interest in undertaking a Legal Judgment Prediction, thus it would be considered as occurring at a pre-case juncture of the legal case lifecycle. This differs by definition from making a prediction once a case has already started, and also implies that there is likely more known about the case once it has started versus when it is in a pre-case mode.

In terms of predicting the outcome of the legal case, it could be that the prediction is aimed at the initial outcome of the case, and for which the case might continue under appeal, gradually and ultimately coming to a final outcome. Thus, it can be ambiguous to refer to a case outcome without clarifying which such outcome the prediction is aiming to derive.

Refer next to **Figure B-7**.

It is often assumed that the start of the case is the prediction deriving juncture, and the target of the prediction is the legal case outcome. Though this is likely a popular choice of the junctures, it is not necessarily the only manner of choice available. In short, at any juncture in the legal case lifecycle, a Legal Judgment Prediction might be rendered.

Thus, a legal case might already be underway and there is interest in predicting the initial case outcome.

Note that this points out too that a target aim can be an intermediate state of the legal case and not solely the end state of the legal case.

For the convenience of reference, Figure B-7 indicates each of the primary junctures as "Jx" whereby the "x" is indicative of a simple numbering of the junctures. This is not meant to be definitive and merely showcases that a numbering scheme can be used to label the stages or phases of a legal case throughout its lifecycle. A legal case could be subdivided into any number Z of junctures (rather than the seven indicated in this example), and thus at any juncture, a prediction can seek to be rendered for some Z+y number of junctures ahead in the lifecycle.

Presumably, the tighter the range between the juncture of the prediction and the juncture of the target stage, the more accurate the prediction can be, assuming all else is equal therein, and likewise that the further along the case is in the lifecycle and thus the higher at which the juncture Z is utilized as a point at which to render a prediction the more accurate the prediction can be, all else being equal.

As mentioned in Section 4 about future research, these postulations about the points at which Legal Judgment Prediction is rendered and targeted are deserving of further research and empirical study.

**3.5 Aims of the Prediction Targeting**

Much of the research literature depicted in Section 1 has been varied as to what the nature of the legal case outcome is to consist of. In other words, it could be that the outcome is solely about whether the case has been ruled in favor or opposition. And though that is obviously a vital element for an outcome, it is not the only such element, and thus it is useful to consider the other elements that might also be considered as prediction targets.

Refer to **Figure B-8**.

As indicated, the legal case outcome can consist of a variety of aspects, including but limited to:
- Prevailing Party
- Mistrial
- Settlement
- Case Dropped
- Verdict
- Penalty
- Etc.

The use of Legal Judgment Prediction can consist of aiming at only one of those such outcomes, or more than one, and/or a combination of those potential outcomes. Research on the nature of LJP can better be utilized by clarifying what the outcome(s) are that are being targeted and also the differences, if any, in how to undertake LJP depending upon the outcome(s) being targeted (this is a recommendation also made in Section 4 on further research).

There are additional targets to be considered.

One is the legal case rationale, which consists of why the outcome was ascertained, or at least a predicted rationale (whether it was the true basis is another question altogether, of course).

Legal Judgment Prediction can also be called upon to predict the legal case costs, the legal case effort, the legal case rationale, and so on.

And, as mentioned in Section 3.4, keep in mind that these predictive matters can occur at any given juncture Z of the legal case lifecycle and be targeting some Z+y juncture, and does not suggest that it is always and only at the start and nor always and only targeting the final state.

**3.6 LJP And Basket Or Bundling of Legal Cases**

Generally, most of the research literature indicated in Section 1 has primarily focused on using Legal Judgment Prediction for taking a specific legal case and predicting the outcome of that specific case.

Legal Judgment Prediction can also be utilized in a basket or bundling manner.

Refer to **Figure B-9**.

As shown, a set of legal cases might be clumped together in a basket of legal cases, or considered a bundle of legal cases, and as a set be used to predict a legal outcome that is anticipated to be the singular outcome applicable to each member of the set.

Thus, there is one singular outcome, and for which it applies presumably to each of the legal cases of the devised set.

Doing this kind of basket or bundling has its own complexities and facets, thus, it is worthwhile to highlight that this is another form in which Legal Judgment Prediction can be utilized, and ergo can be studied as a variant thereof.

### 3.7 Additional Crucial Facets

To further aid in identifying facets that are crucial to the maturity of Legal Judgement Prediction, consider the elements indicated in **Figure B-10**.

The elements shown include:
- Transparency
- Actor Predictors
- Biases Handling
- Probabilistic Tendencies

Consider each of these elements.

- Transparency.

In brief, one consideration involves the transparency involved in a Legal Judgment Prediction. The method used for the LJP might be available for inspection and fully transparent, or it might be a Black Box that is considered opaque. Another facet is whether the LJP can explain how it worked, or whether it might be inscrutable or lacking in being able to be interpreted.

- Actor Predictors

In brief, the question arises as to how the Legal Judgment Prediction will be carried out. This can consist of a human-only based prediction, an AI-only based prediction, or a human and AI jointly undertaken prediction.

- Biases Handling

In brief, there are many aspects to be addressed regarding the potential for biases being an ingredient within a Legal Judgment Prediction. There can be hidden biases, which can distort or alter the nature of the prediction and the predictive capacity. There can be explicitly noted biases, which then need to be considered in light of the predictions rendered. Various bias correction methods and approaches can be utilized. And so on.

- Probabilistic Tendencies

Much of the research literature as depicted in Section 1 does not especially bring forth the aspect that the prediction is undoubtedly one of a probabilistic nature. In a sense, without explicitly mentioning the certainty or uncertainty of a prediction, the default tends to imply that the certainty is complete, though this is rarely the likely case. As such, it is vital to acknowledge and recognize the uncertainties involved and then indicate what those consist of, along with how the certainty can be impacted by the methods used in the course of the Legal Judgment Prediction.

### 3.8 AILR and LJP Impact

An important consideration worth exploring is the over-time judicial reliance upon human judicial judgment versus AILR capabilities.

Refer to **Figure B-11**.

A graph is portrayed that consists of the magnitude of judicial reliance on the Y-axis (ranging from low to high), while time is exhibited on the X-axis. Time is indicated as beginning at t-0 and proceeding forward in time, of which there are two especially salient points in time marked, consisting of time t-r and time t-u, explained momentarily.

A dashed line representing the reliance upon human judicial judgment is shown as starting at the high position on the Y-axis and then proceeding in a sloping downward motion over-time, inexorably proceeding toward the lower ends of reliance.

Meanwhile, a second line is shown as starting at the origin point of zero, proceeding on an upward slope over-time, and indicative of the AILR capabilities that will presumably increase over time and thus it is assumed there will be an increasing reliance upon the AILR for participating in or providing judicial judgment.

Note that the graph is not drawn to scale and is intended merely as an overall gauge or macroscopic illustration of the issues being discussed herein.

At some point in time, designated as t-r, the reliance upon AILR begins to exceed the reliance upon human judicial judgment (so postulated).

It is not a foregone conclusion that this will necessarily occur and purely indicated as worthwhile for discussion. Likewise, it is not necessarily the case that the rising AILR reliance necessitates a reducing reliance on human judicial judgment, though this is the postulated theory that oft is argued about the gradual advance of AILR.

After passing the crossover, there is a Zone S, into which it is possible that the problematic aspects of performing Legal Judgment Prediction might wane, due to the facet that the AILR is becoming dominant in the reliance upon rendering judicial decisions and (if one assumes) the AILR will be explicative and thus presumably transparent. This transparency would suggest that predicting what the AILR is going to judicial proffer is much more predictable and straightforward (though, not necessarily the case if the AILR has been allowed to become opaque or otherwise closed in its inner workings).

The line representing the reliance upon human judicial judgment does not touch the X-axis as it is unknown as to whether there would ever be a point at which all human judicial judgment would be no longer utilized such that only AILR was utilized (this is a classic open debate).

**4 Additional Considerations and Future Research**

As earlier indicated, Legal Judgment Prediction (LJP) is a longstanding and open issue in the theory and practice-of-law. Predicting the nature and outcomes of judicial matters is important. Intrinsically, generating judicially laden predictions has been limited in utility and exactitude, as illustrated via the literature review of Section 1. Though various methods and techniques to predict legal cases and judicial actions have emerged over time, especially arising via the advent of computer-based modeling, including simple calculative methods to highly sophisticated and complex statistical models, much still needs to be done to improve LJP.

Artificial Intelligence (AI) based approaches have been increasingly utilized and will undoubtedly have a pronounced impact on how LJP is performed and its predictive accuracy. Per Section 3, Legal Judgment Prediction can holistically be understood by using the Levels of Autonomy (LoA) of AI Legal Reasoning, plus, other considerations are further to be advanced, including LJP probabilistic tendencies, biases handling, actor predictors, transparency, judicial reliance, legal case outcomes, and other crucial elements entailing the overarching legal judicial milieu.

Future research is needed to explore in greater detail the manner and means by which AI-enablement will occur in the law along with the potential for both positive and adverse consequences in LJP. Autonomous AILR is likely to materially impact the effort, theory, and practice of Legal Judgment Prediction, including as a minimum playing a notable or possibly even pivotal role in such advancements.

**About the Author**

Dr. Lance Eliot is the Chief AI Scientist at Techbrium Inc. and a Stanford Fellow at Stanford University in the CodeX: Center for Legal Informatics. He previously was a professor at the University of Southern California (USC) where he headed a multi-disciplinary and pioneering AI research lab. Dr. Eliot is globally recognized for his expertise in AI and is the author of highly ranked AI books and columns.

42. Hage, Jaap (2000). "Dialectical Models in Artificial Intelligence and Law," Artificial Intelligence and Law.

43. Hall, Kermit, and William Wiecek, Paul Finkelman (1996). American Legal History: Cases and Materials, 2nd ed. Oxford University Press.

44. Hall, Mark, and Ronald Wright (2006). "Systematic Content Analysis of Judicial Opinions," Volume 96, Issue 1, California Law Review.

45. Harbert, T. (2013). "The Law Machine," Volume 50, Issue 11, IEEE Spectrum.

46. Hart, H (1961). The Concept of Law. Oxford University Press.

47. Holmes, Oliver Wendell (1897). "The Path of the Law," Harvard Law Review.

48. Huhn, Wilson (2000). "Teaching Legal Analysis Using a Pluralistic Model of Law," Volume 36, Number 3, Gonzaga Law Review.

49. Huhn, Wilson (2003). "The Stages of Legal Reasoning: Formalism, Analogy, and Realism," Volume 48, Villanova Law Review.

50. Kaplow, Louis (1992). "Rules Versus Standards: An Economic Analysis," Volume 42, Duke Law Journal.

51. Kastellec, Jonathan (2010). "The Statistical Analysis of Judicial Decisions and Legal Rules with Classification Trees," Volume 7, Number 2, Journal of Empirical Legal Studies.

52. Katz, Daniel, and Michael Bommarito, Josh Blackman (2017). "A General Approach for Predicting the Behavior of the Supreme Court of the United States," Volume 12, Number 4, Plos One.

53. Keown, R. (1980). "Mathematical Models for Legal Prediction," Volume 2, Issue 1, Computer/Law Journal.

54. Kort, F. (1957). "Predicting Supreme Court Decisions Mathematically: A Quantitative Analysis of the "Right to Counsel" Cases," Volume 51, Number 1, American Political Science Review.

55. Lauderdale, Benjamin, and Tom Clark (2012). "The Supreme Court's Many Median Justices," Volume 106, Number 4, American Political Science Review.

56. Lawlor, Reed (1963). "What Computers Can Do: Analysis and Prediction of Judicial Decisions," Volume 49, Number 4, American Bar Association Journal.

57. Llewellyn, Karl (1950). "Remarks on the Theory of Appellate Decision and the Rules or Canons About How Statutes Are to Be Construed," Volume 3, Number 4, Vanderbilt Law Review.

58. Llewellyn, Karl (1960). The Common Law Tradition: Deciding Appeals. Quid Pro Books.

59. Luo, Bingfeng, and Yansong Feng, Jianbo Xu, Xiang Zhang, Dongyan Zhao (2017). "Learning to Predict Charges for Criminal Cases with Legal Basis," arXiv preprint arXiv:1707.09168.

60. MacCormick, Neil (1978). Legal Reasoning and Legal Theory.

61. Markou, Christopher, and Simon Deakin (2020). "Is Law Computable? From Rule of Law to Legal Singularity," May 4, 2020, SSRN, University of Cambridge Faculty of Law Research Paper.
24

62. Martin, Andrew, and Kevin Quinn, Theodore Ruger, Pauline Kim (2004). "Competing Approaches to Predicting Supreme Court Decision Making," Perspectives on Politics.

63. McCarty, L. (1995)." An implementation of Eisner v. Macomber," Proceedings of the 5th International Conference on Artificial Intelligence and Law.

64. McCarty, Thorne (1977). "Reflections on TAXMAN: An Experiment in Artificial Intelligence and Legal Reasoning," January 1977, Harvard Law Review.

65. McGinnis, John, and Russell G. Pearce (2014). "The Great Disruption: How Machine Intelligence Will Transform the Role of Lawyers in the Delivery of Legal Services," Volume 82, Number 6, Fordham Law Review.

66. McGinnis, John, and Steven Wasick (2015). "Law's Algorithm," Volume 66, Florida Law Review.

67. Mnookin, Robert, and Lewis Kornhauser (1979). "Bargaining in the Shadow of the Law," Volume 88, Number 5, April 1979, The Yale Law Review.

68. Nagel, Stuart (1963). "Applying Correlation Analysis to Case Prediction," Volume 42, Texas Law Review.

69. Noonan, John (1961). "Review of Hart's Book 'The Concept of Law'," Volume 7, Issue 1, The American Journal of Jurisprudence.

70. Prakken, Henry (1995). "From Logic to Dialectics in Legal Argument," Proceedings of the 5th International Conference on Artificial Intelligence and Law.

71. Reinbold, Patric (2020). "Taking Artificial Intelligence Beyond the Turing Test," Volume 20, Wisconsin Law Review.

72. Remus, Dana, and Frank Levy, "Can Robots be Lawyers? Computers, Robots, and the Practice of Law," Volume 30, Georgetown Journal of Legal Ethics.

73. Rich, Michael (2016). "Machine Learning, Automated Suspicion Algorithms, and the Fourth Amendment," Volume 164, University of Pennsylvania Law Review.

74. Rissland, Edwina (1990). "Artificial Intelligence and Law: Stepping Stones to a Model of Legal Reasoning," Yale Law Journal.

75. SAE (2018). Taxonomy and Definitions for Terms Related to Driving Automation Systems for On-Road Motor Vehicles, J3016-201806, SAE International.

76. Schauer, Frederick (1998). "Prediction and Particularity," Volume 78, Boston University Law Review.

77. Segal, Jeffrey (1984). "Predicting Supreme Court Cases Probabilistically: The Search and Seizure Cases," Volume 78, Issue 4, American Political Science Review.

78. Shaikh, Rafe Athar and Tirath Prasad Sahu, Veena Anand (2020). "Predicting Outcomes of Legal Cases based on Legal Factors using Classifiers," ScienceDirect.

79. Sunstein, Cass (2001). "Of Artificial Intelligence and Legal Reasoning," University of Chicago Law School, Working Papers.

80. Sunstein, Cass, and Kevin Ashley, Karl Branting, Howard Margolis (2001). "Legal Reasoning and Artificial Intelligence: How Computers 'Think' Like Lawyers," Symposium: Legal Reasoning and Artificial Intelligence, University of Chicago Law School Roundtable.
25

**Figure A-1**

### AI & Law: Levels of Autonomy For AI Legal Reasoning (AILR)

| Level | Descriptor | Examples | Automation | Status |
|---|---|---|---|---|
| 0 | No Automation | Manual, paper-based (no automation) | None | De Facto - In Use |
| 1 | Simple Assistance Automation | Word Processing, XLS, online legal docs, etc. | Legal Assist | Widely In Use |
| 2 | Advanced Assistance Automation | Query-style NLP, ML for case prediction, etc. | Legal Assist | Some In Use |
| 3 | Semi-Autonomous Automation | KBS & ML/DL for legal reasoning & analysis, etc. | Legal Assist | Primarily Prototypes & Research Based |
| 4 | AILR Domain Autonomous | Versed only in a specific legal domain | Legal Advisor (law fluent) | None As Yet |
| 5 | AILR Fully Autonomous | Versatile within and across all legal domains | Legal Advisor (law fluent) | None As Yet |
| 6 | AILR Superhuman Autonomous | Exceeds human-based legal reasoning | Supra Legal Advisor | Indeterminate |

*Figure 1: AI & Law - Autonomous Levels by Rows*          *Source Author: Dr. Lance B. Eliot*   V1.3

**Figure A-2**

| | Level 0 | Level 1 | Level 2 | Level 3 | Level 4 | Level 5 | Level 6 |
|---|---|---|---|---|---|---|---|
| | | | AI & Law: Levels of Autonomy For AI Legal Reasoning (AILR) | | | | |
| **Descriptor** | No Automation | Simple Assistance Automation | Advanced Assistance Automation | Semi-Autonomous Automation | AILR Domain Autonomous | AILR Fully Autonomous | AILR Superhuman Autonomous |
| **Examples** | Manual, paper-based (no automation) | Word Processing, XLS, online legal docs, etc. | Query-style NLP, ML for case prediction, etc. | KBS & ML/DL for legal reasoning & analysis, etc. | Versed only in a specific legal domain | Versatile within and across all legal domains | Exceeds human-based legal reasoning |
| **Automation** | None | Legal Assist | Legal Assist | Legal Assist | Legal Advisor (law fluent) | Legal Advisor (law fluent) | Supra Legal Advisor |
| **Status** | De Facto – In Use | Widely In Use | Some In Use | Primarily Prototypes & Research-based | None As Yet | None As Yet | Indeterminate |

*Figure 2: AI & Law - Autonomous Levels by Columns*     *Source Author: Dr. Lance B. Eliot*

V1.3

**Figure B-1**

### Legal Judgment Prediction: Levels of Autonomy For AI Legal Reasoning (AILR)

| | Level 0 | Level 1 | Level 2 | Level 3 | Level 4 | Level 5 | Level 6 |
|---|---|---|---|---|---|---|---|
| **Descriptor** | No Automation | Simple Assistance Automation | Advanced Assistance Automation | Semi-Autonomous Automation | AILR Domain Autonomous | AILR Fully Autonomous | AILR Superhuman Autonomous |
| **Examples** | Manual, paper-based (no automation) | Word Processing, XLS, online legal docs, etc. | Query-style NLP, ML for case prediction, etc. | KBS & ML/DL for legal reasoning & analysis, etc. | Versed only in a specific legal domain | Versatile within and across all legal domains | Exceeds human-based legal reasoning |
| **Automation** | None | Legal Assist | Legal Assist | Legal Assist | Legal Advisor (law fluent) | Legal Advisor (law fluent) | Supra Legal Advisor |
| **Status** | De Facto – In Use | Widely In Use | Some In Use | Primarily Prototypes & Research-based | None As Yet | None As Yet | Indeterminate |
| **AI-Enabled Legal Judgment Prediction** | n/a | Rudimentary Calculative | Complex Statistical | Symbolic Intermixed | Domain Predictive | Holistic Predictive | Pansophic Predictive |

*Figure 1: Legal Judgment Prediction (LJP) - Autonomous Levels of AILR by Columns*  *Source Author: Dr. Lance B. Eliot*  V1.3

**Figure B-2**

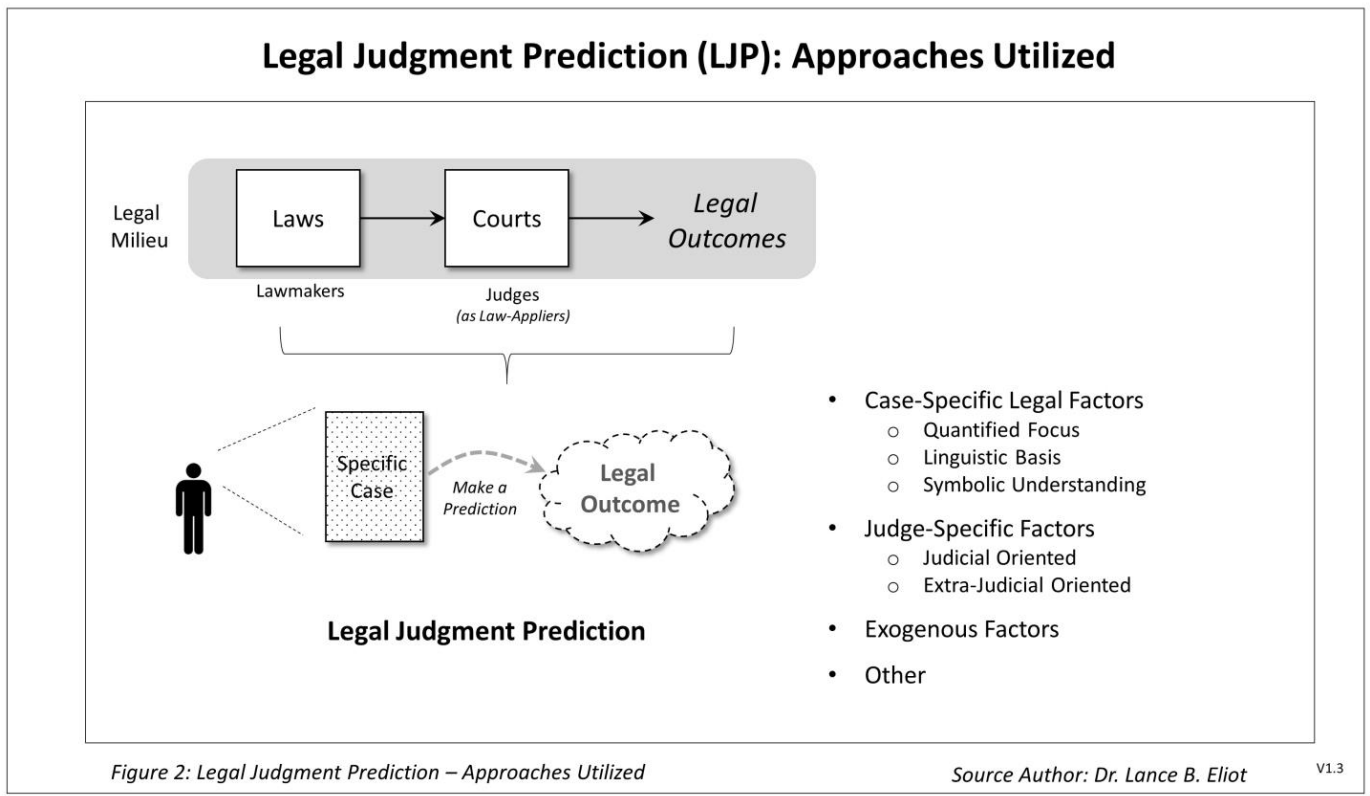

Figure 2: Legal Judgment Prediction – Approaches Utilized

**Figure B-3**

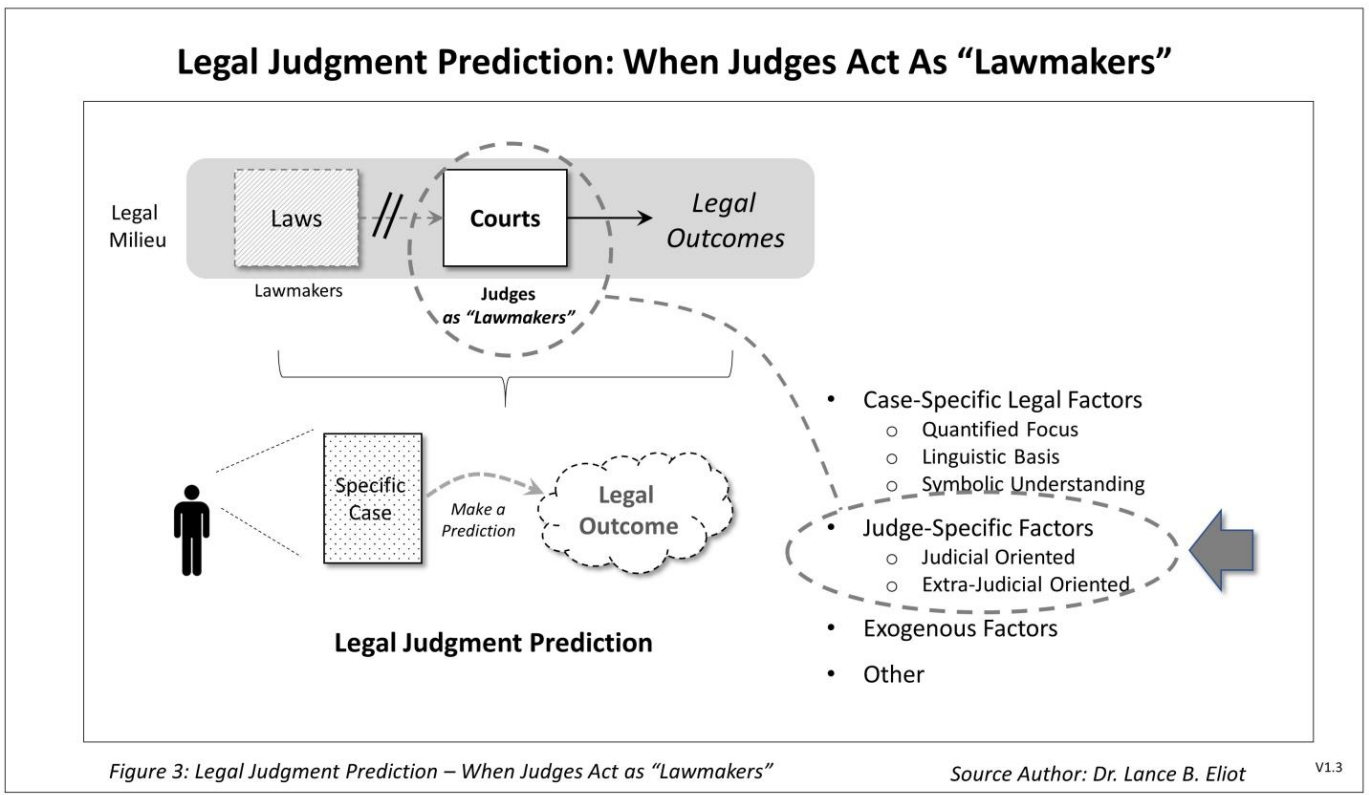

**Figure B-4**

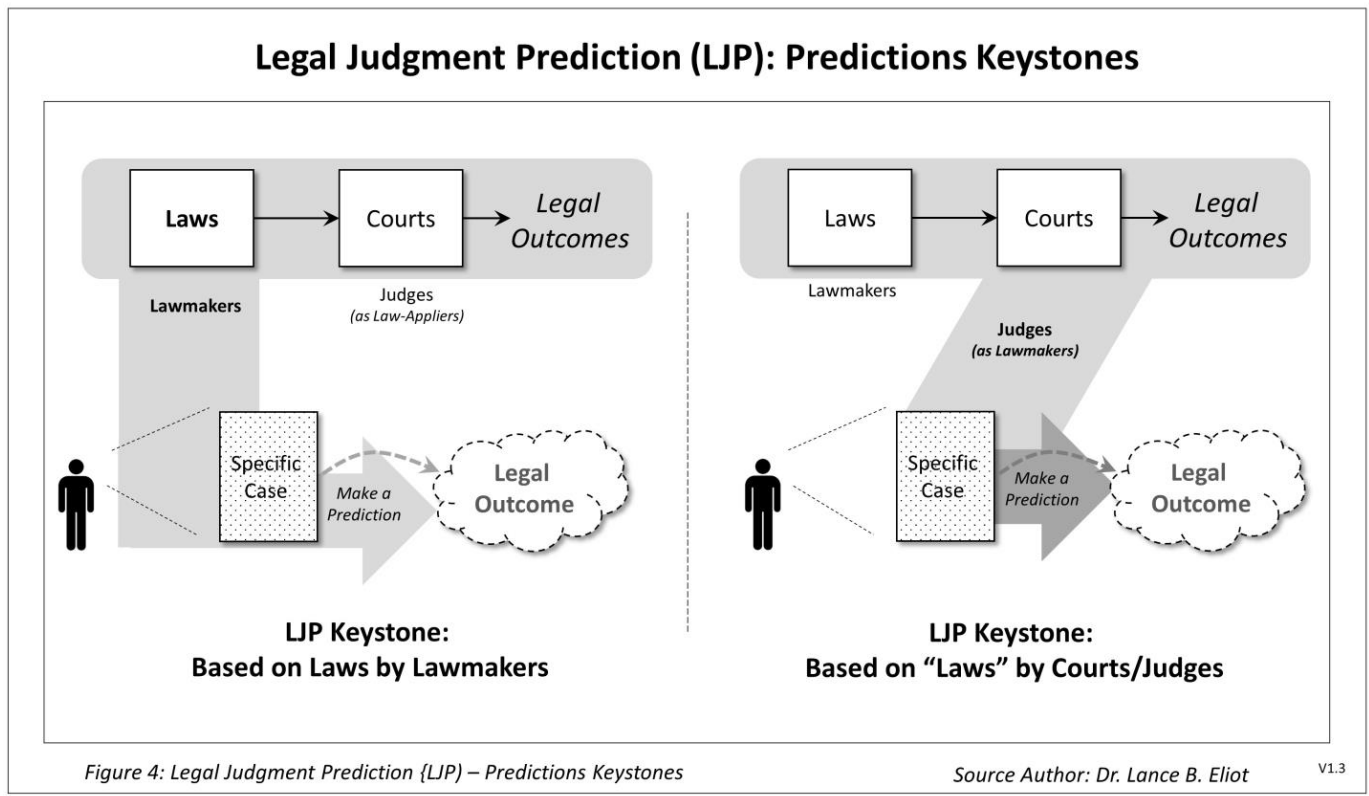

*Figure 4: Legal Judgment Prediction {LJP} – Predictions Keystones*   Source Author: Dr. Lance B. Eliot   V1.3

**Figure B-5**

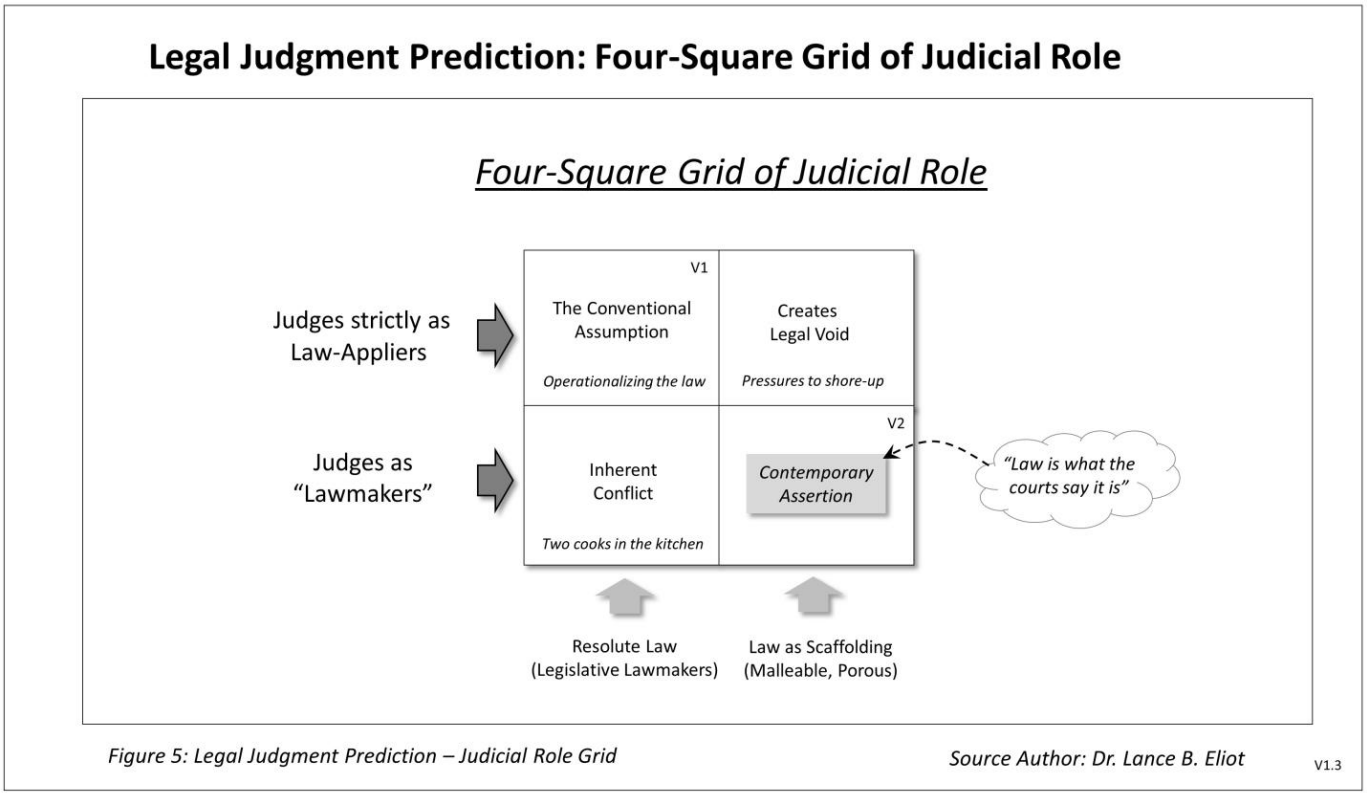

Figure 5: Legal Judgment Prediction – Judicial Role Grid

**Figure B-6**

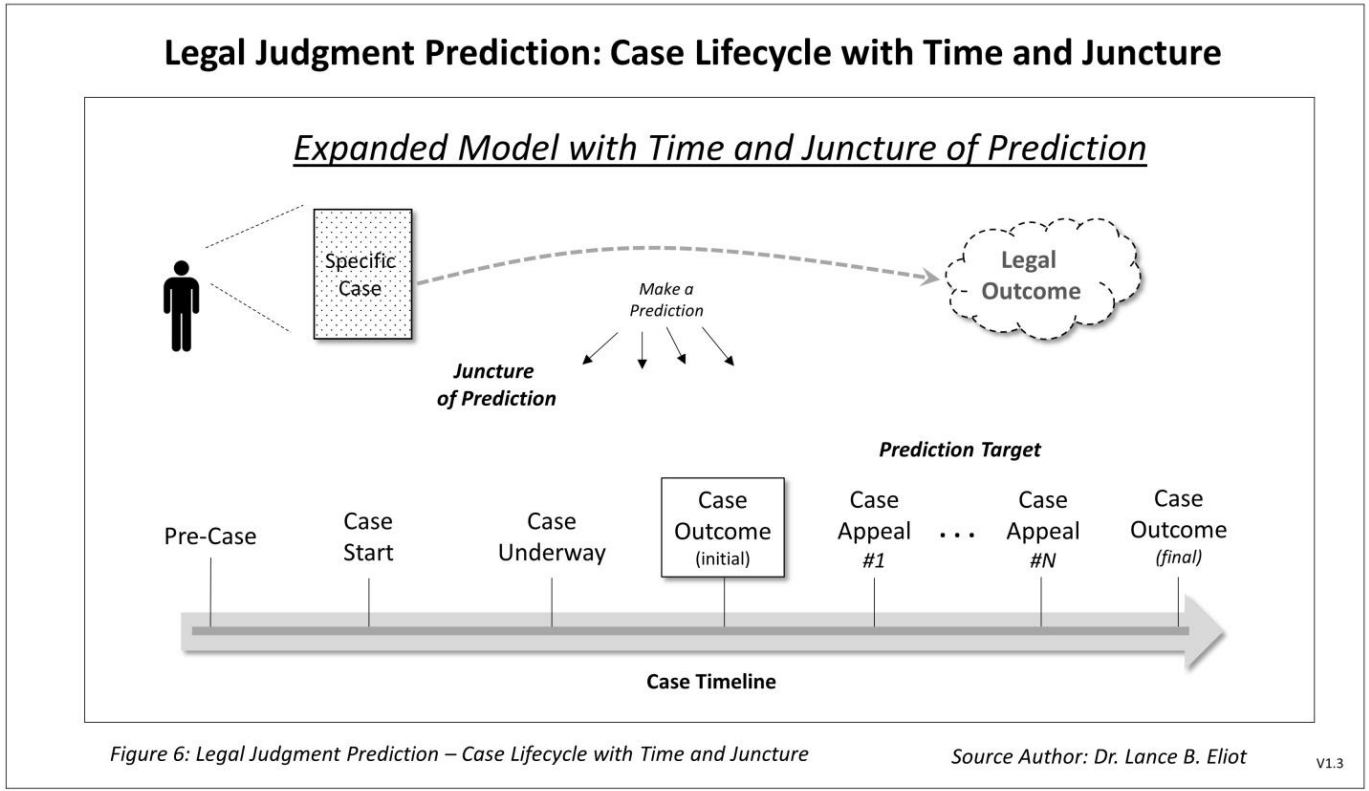

**Figure B-7**

## Legal Judgment Prediction: Lifecycle with Prediction Time/Juncture

| | From/To | Pre-Case | Case Start | Case Underway | Case Outcome (initial) | Case Appeal #1 | Case Appeal #N | Case Outcome (final) |
|---|---|---|---|---|---|---|---|---|
| J1 | Pre-Case | | J2 | J3 | J4 | J5 | J6 | J7 |
| J2 | Case Start | | | J3 | J4 | J5 | J6 | J7 |
| J3 | Case Underway | | | | J4 | J5 | J6 | J7 |
| J4 | Case Outcome (initial) | | | | | J5 | J6 | J7 |
| J5 | Case Appeal #1 | | | | | | J6 | J7 |
| J6 | Case Appeal #N | | | | | | | J7 |
| J7 | Case Outcome (final) | | | | | | | |

*Figure 7: LJP Chart of Case Lifecycle with Prediction Time/Juncture*   *Source Author: Dr. Lance B. Eliot*   V1.3

**Figure B-8**

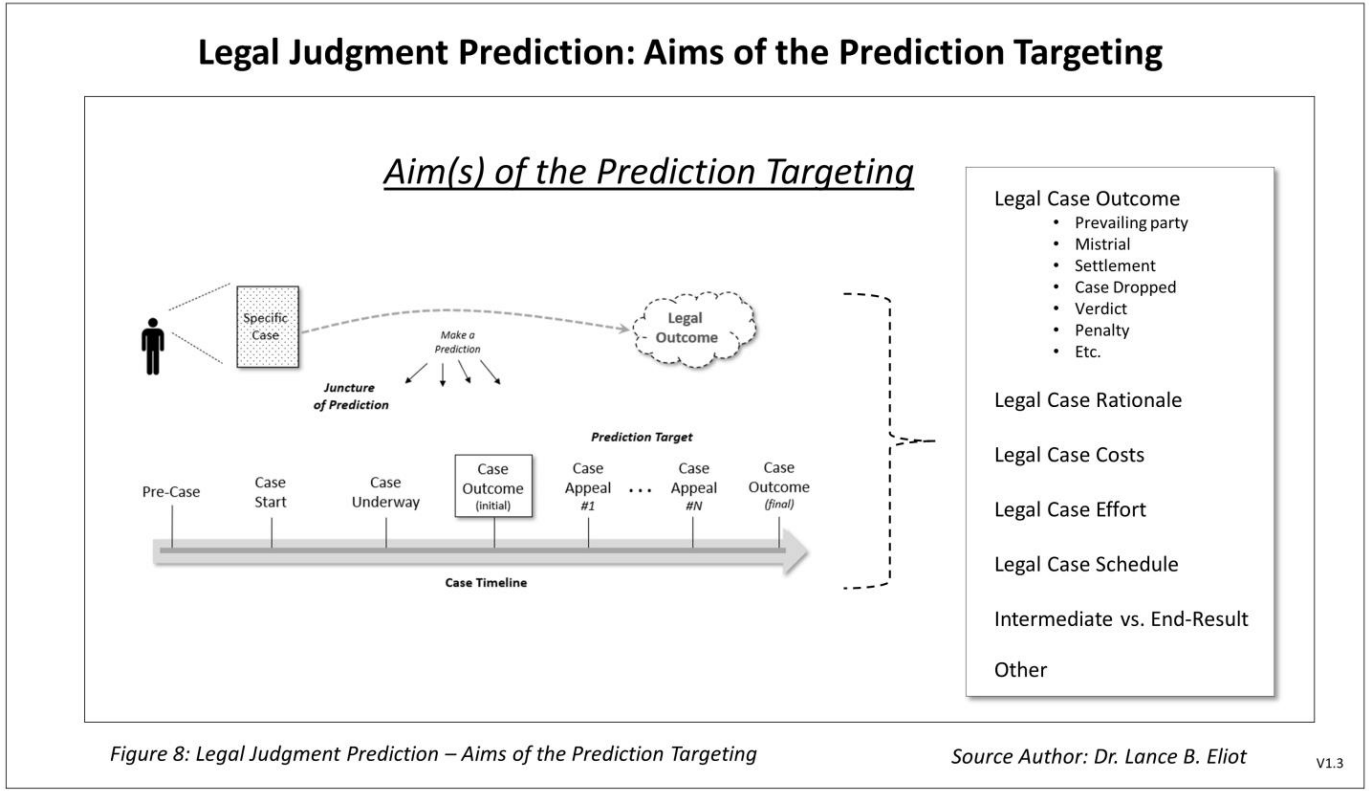

Figure 8: Legal Judgment Prediction – Aims of the Prediction Targeting

**Figure B-9**

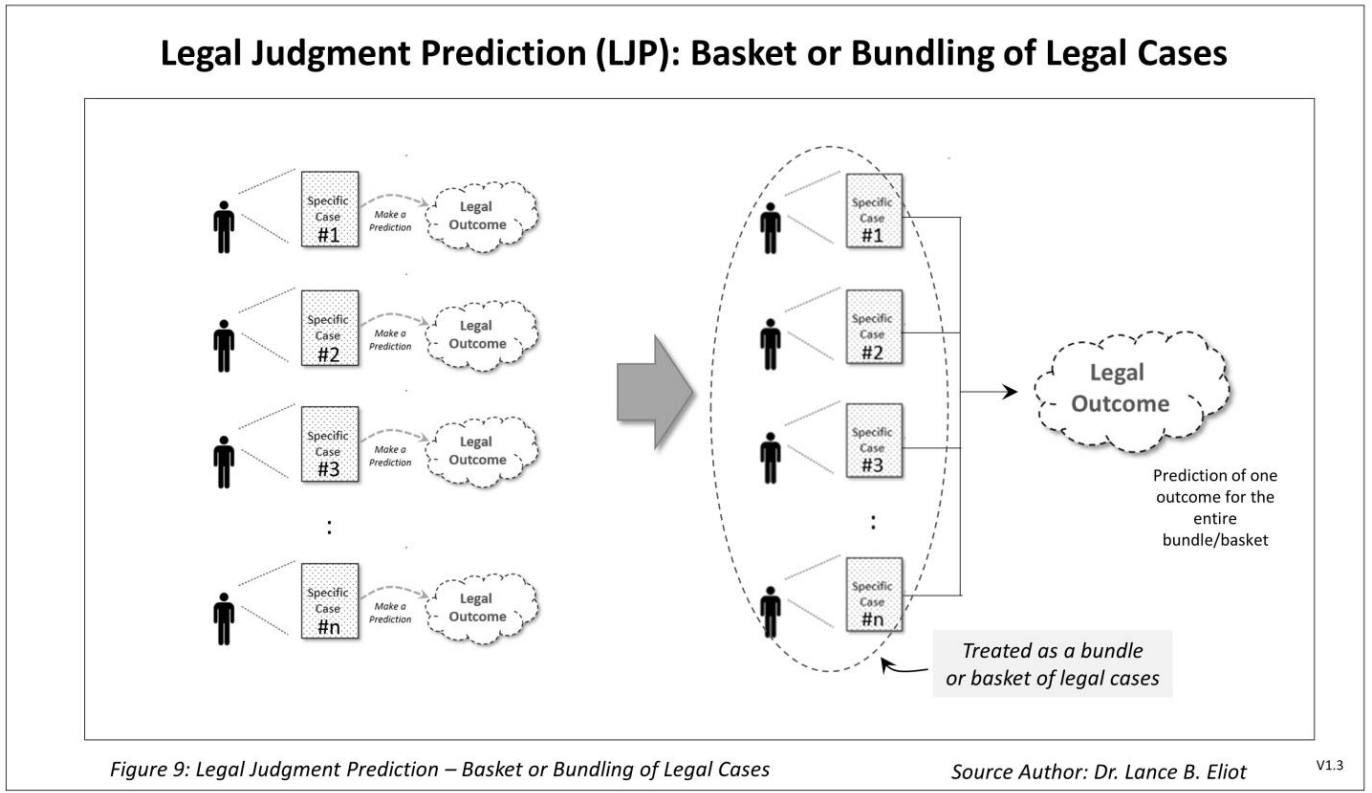

Figure 9: Legal Judgment Prediction – Basket or Bundling of Legal Cases

**Figure B-10**

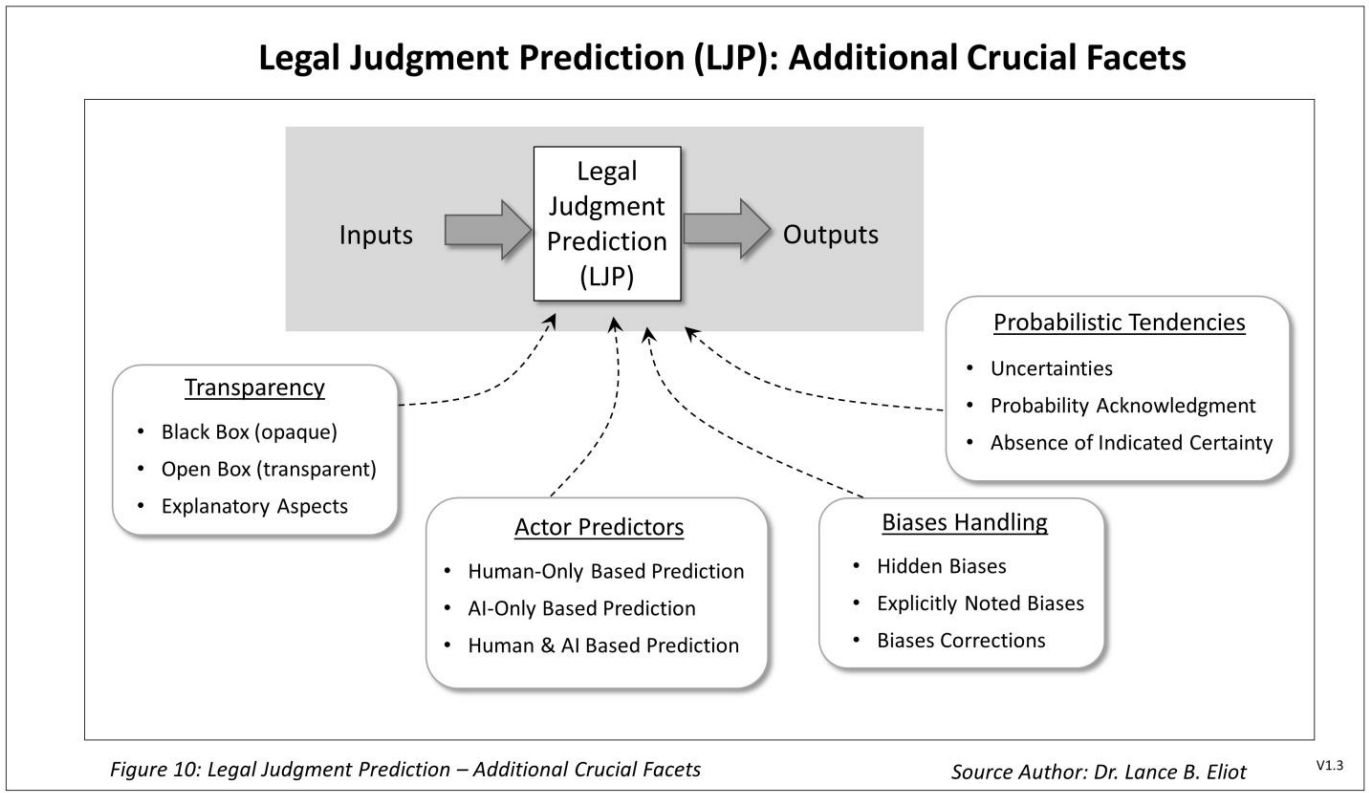

Figure 10: Legal Judgment Prediction – Additional Crucial Facets

**Figure B-11**

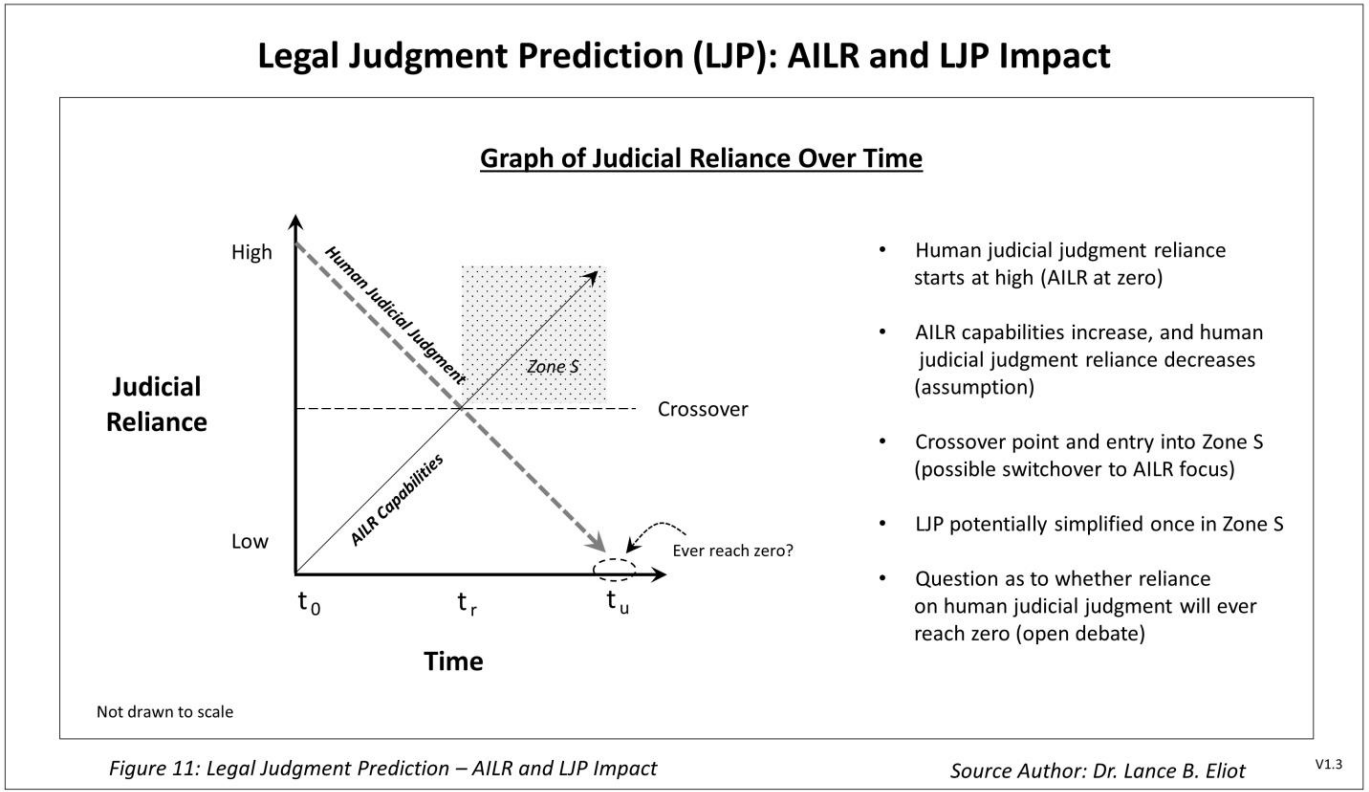

Figure 11: Legal Judgment Prediction – AILR and LJP Impact